\documentclass[12pt,preprint,url]{aastex}
\usepackage{epsfig}
\usepackage{graphicx} 
\usepackage{amsbsy}
\usepackage{amsmath}
\newcommand{\be}{\begin{equation}}
\newcommand{\ee}{\end{equation}}

\def\ltsima{$\; \buildrel < \over \sim \;$}

\def\lsim{\lower.5ex\hbox{\ltsima}}
\def\gtsima{$\; \buildrel > \over \sim \;$}
\def\gsim{\lower.5ex\hbox{\gtsima}}

\begin{document}
\title{Avalanche dynamics of radio pulsar glitches}


\author{A. Melatos\altaffilmark{1}, C. Peralta\altaffilmark{1,}\altaffilmark{2}
and J. S. B. Wyithe\altaffilmark{1} }

\email{a.melatos@unimelb.edu.au}

\altaffiltext{1}{School of Physics, University of Melbourne,
Parkville, VIC 3010, Australia}

\altaffiltext{2}{Max-Planck-Institut f\"ur
Gravitationsphysik, Albert-Einstein-Institut, Am M\"uhlenberg 1,
D-14476 Golm, Germany}

\begin{abstract}
\noindent
We test statistically the hypothesis that radio pulsar glitches result from
an avalanche process, in which angular momentum is transferred erratically
from the flywheel-like superfluid in the star to the slowly decelerating, solid crust
via spatially connected chains of local, impulsive, threshold-activated events,
so that the system fluctuates around a self-organised critical state.
Analysis of the glitch population (currently 285 events from 101 pulsars)
demonstrates that the size distribution in individual pulsars
is consistent with being scale invariant, as expected for an avalanche process.
The measured power-law exponents fall in the range $-0.13\leq a \leq 2.4$,
with $a\approx 1.2$ for the youngest pulsars.
The waiting-time distribution is consistent with being 
exponential in seven out of nine pulsars
where it can be measured reliably,
after adjusting for observational limits on the minimum waiting time,
as for a constant-rate Poisson process.
PSR J0537$-$6910 and PSR J0835$-$4510 are the exceptions;
their waiting-time distributions show evidence of quasiperiodicity.
In each object, stationarity requires that the rate $\lambda$ equals
$- \epsilon \dot{\nu} / \langle\Delta\nu\rangle$,
where $\dot{\nu}$ is the angular acceleration of the crust,
$\langle\Delta\nu\rangle$ is the mean glitch size,
and $\epsilon\dot{\nu}$ is the relative angular acceleration of the crust and superfluid.
Measurements yield $\epsilon \leq 7 \times 10^{-5}$
for PSR J0358$+$5413 and $\epsilon \leq 1$ (trivially)
for the other eight objects, which have $a < 2$.
There is no evidence that $\lambda$ changes monotonically
with spin-down age.
The rate distribution itself is fitted reasonably well by an exponential
for $\lambda \geq 0.25\,{\rm yr^{-1}}$,
with $\langle \lambda \rangle = 1.3^{+0.7}_{-0.6}\,{\rm yr^{-1}}$.
For $\lambda < 0.25\,{\rm yr^{-1}}$, its exact form is unknown;
the exponential overestimates the number of glitching pulsars
observed at low $\lambda$, where the limited total observation time
exercises a selection bias.
In order to reproduce the aggregate waiting-time distribution of the
glitch population as a whole,
the fraction of pulsars with $\lambda > 0.25\,{\rm yr^{-1}}$
must exceed $\sim 70$ per cent.
\end{abstract}

\keywords{dense matter --- pulsars: general --- stars: interior --- stars: neutron --- stars: rotation}

\section{Introduction 
 \label{sec:gli1}}
Glitches are tiny, impulsive, randomly timed increases in the spin frequency $\nu$ 
of a rotation-powered pulsar,
sometimes accompanied by an impulsive change in the frequency derivative $\dot{\nu}$.
They are to be distinguished from timing noise,
a type of rotational irregularity
where pulse arrival times wander continuously,
although there is evidence that timing noise is the cumulative result of 
frequent microglitches in certain pulsars
\citep{cor85,dal95}.

At the time of writing,
285 glitches in total have been detected in 101 objects
($\sim 6\%$ of the known radio pulsar population),
the majority in the last four years, facilitated by the
Parkes Multibeam Survey,
refined multifrequency ephemerides,
and better interference rejection algorithms
\citep{hob02,kra03,kra05,lew05,jan06}.
Efforts to analyse the data statistically have focused on
the correlation of glitch activity with age
\citep{mck90,she96,ura99,lyn00,wan00}
and Reynolds number
\citep{peralta06,mel07},
the post-glitch relaxation time-scale
\citep{wan00,won01},
the size distribution
\citep{mor93a,mor93b,peralta06},
and the correlation between glitch sizes and waiting times
\citep{wan00,won01,mid06,peralta06}.
\citet{hob02} reviewed the role of observational selection effects.

Most glitching pulsars ($65\%$) have been seen to glitch once,
but a minority glitch repeatedly;
the current record holder is PSR J1740$-$3015, with 33 glitches.
Of those objects which glitch repeatedly,
most do so at unpredictable intervals, 
but two (PSR J0537$-$6910 and Vela) are quasiperiodic;
Vela, in particular, has been likened to a relaxation oscillator
\citep{lyn96}.
The fractional increase in $\nu$ spans seven decades 
($3\times 10^{-11} \leq \Delta\nu \leq 2\times 10^{-4}$) 
across the glitch population 
and as many as four decades in a single object
(e.g.\ $7\times 10^{-10} \leq \Delta\nu \leq 2\times 10^{-6}$
in PSR J1740$-$3015).
The spin-down age $\tau_{\rm c}= -\nu/(2\dot{\nu})$
of glitching pulsars spans four decades,
from $1\times 10^3\,{\rm yr}$ to $3\times 10^7\,{\rm yr}$.
In many respects, therefore, the glitch phenomenon is {\em scale invariant}.
This striking property invites physical interpretation.

Theories of pulsar glitches have focused mainly on the local microphysics
of the superfluid in the stellar interior and its coupling to the solid crust,
for example the strength of vortex pinning
\citep{and75,jon98}, 
the rate of vortex creep
\citep{lin96},
or the conditions for exciting superfluid turbulence
\citep{per05,per06,mel07,and07}.
Ultimately, however, the local microphysics must be synthesized with
the global, {\em collective} dynamics in order to make full contact 
with observational data.
(Likewise, a practical model of earthquakes must synthesize
the microphysics of rock fracture with the
macrodynamics of interacting tectonic plates.)
For example, if approximately $10^{16} (\Delta\nu/1\,{\rm Hz})$ 
vortices unpin from crustal lattice sites in sympathy during a glitch,
they must communicate rapidly across distances much greater than 
their separation. How?
And why does the number that unpin fluctuate so dramatically
(by up to four orders of magnitude)
from glitch to glitch in a single pulsar,
while always amounting to a small fraction ($\Delta\nu/\nu$) 
of the total?

Such collective, scale invariant behavior is a generic feature of 
a class of natural and synthetic far-from-equilibrium systems, 
called self-organized critical systems, 
that are discrete, interaction dominated, and slowly driven, 
and that adjust internally via erratic, spatially connected {\em avalanches}
of local, impulsive, threshold-activated, relaxation events
\citep{jen98}.
Such systems fluctuate around a stationary state
towards which they evolve spontaneously,
in which global driving balances local relaxation on average over the long term.
The archetype of a self-organized critical system is the sandpile
\citep{bak87}.

In this paper, we study pulsar glitches as an avalanche process, 
as first proposed by \citet{mor93a}.
After reviewing self-organised criticality in \S\ref{sec:gli2},
we define the statistical sample on which our study is based
(\S\ref{sec:gli3})
and analyze the observed distribution of glitch sizes
(\S\ref{sec:gli4})
and waiting times
(\S\ref{sec:gli5}).
Some implications for glitch physics are explored in \S\ref{sec:gli6}.
We only include radio pulsars in the sample, to preserve its homogeneity,
even though glitches have now been observed in anomalous X-ray pulsars
(magnetars) as well
\citep{dal03,kas03}.

\section{Avalanche dynamics
 \label{sec:gli2}}
A system in a self-organised critical state exhibits the following
distinguishing features
\citep{jen98}.
\begin{enumerate}
\item
It is composed of many discrete, mutually interacting elements,
whose motions are dominated by local (e.g.\ nearest neighbor) rather than
global (e.g.\ mean field) forces.
\footnote{Tectonic plates, or grains of sand in a pile,
are terrestrial examples of interacting elements.}
\item
Each element moves when the local force exceeds a threshold
(stick-slip motion).
Hence stress accumulates sustainedly at certain random locations
while relaxing quickly elsewhere;
at any instant, the system houses numerous metastable 
stress reservoirs, separated by relaxed zones.
\item
An external force drives the system slowly, in the sense that
elements adjust to local forces rapidly compared to the driver time-scale.
Combined with local thresholds, this ensures that the system
evolves quasistatically through a history-dependent sequence of metastable states
(a huge number of which are available).
\item
Transitions from one metastable state to the next occur
via avalanches:
spatially connected chains of local equilibration events,
in which one element relaxes and
redistributes some local stress to its neighbors,
which in turn can exceed their thresholds and relax
(knock-on effect).
The duration of even the largest avalanches is short compared to
the driving time-scale (see previous point).
\item
Avalanches have no preferred scale: they can involve a few
(commonly) or all (rarely) of the elements in the system.
Their sizes and lifetimes follow power-law distributions,
whose exponents are related.
The numerical values of the exponents depend on the spatial dimensionality 
of the system,
the spatial symmetries of the local forces and redistributive channels,
the strength of the local forces 
\citep{fie95}, 
and the level of conservation 
\citep{ola92}.
\footnote{In this respect, far-from-equilibrium critical systems
differ from equilibrium critical systems 
(e.g.\ second-order phase transition in a ferromagnet),
whose exponents depend only on the dimensionality of the system
and its order parameter(s).}
\item
Over the long term, the system tends to a critical state,
which is stationary on average but not instantaneously.
For example, on average, the power input by the external driver equals the 
energy per unit time released by avalanches.
But there are fluctuations, because, at any instant, 
a random amount of energy is stored in metastable local reservoirs.
\end{enumerate}

Avalanche dynamics are generically observed in nature
when conditions (i)--(iii) are met,
and properties (iv)--(vi) emerge irrespective of the detailed microphysics
\citep{jen98}.
Likewise, in this paper, we remain agnostic about the microphysics
of pulsar glitches;
the statistical analysis presented below makes no assumptions
in this regard.
Nevertheless, it is striking that the traditional glitch paradigm ---
collective unpinning of quantized superfluid vortices interacting with
an inhomogeneous, slowly decelerating crust ---
conforms closely with (i)--(vi)
\citep{and75,alp96}.
So too does an alternative paradigm, based on crust fracture
\citep{alp96,mid06},
whose terrestrial counterpart (plate tectonics)
is renowned as an archetype of self-organized criticality
\citep{sor91}.
We elucidate the analogy briefly before continuing.

Consider a rectilinear array of quantized vortices, 
each carrying circulation $\kappa$,
spaced evenly according to Feynman's rule 
($4\pi\nu/\kappa$ vortices per unit area) in the neutron superfluid
permeating the inner crust of a neutron star.
A small percentage of the vortices are pinned to defects and/or nuclei 
at random locations in the crustal lattice, clustered to varying degrees
\citep{alp96,won01}
As $\nu$ decreases gradually due to electromagnetic spin down,
most vortices move apart and the outermost ones are expelled.
However, the pinned vortices stay (nearly) fixed, 
in metastable reservoirs separated by relaxed zones [see (ii)],
creeping slowly between adjacent pinning sites in response to thermal fluctuations
\citep{lin93}.
The reservoirs are identical to the capacitive elements
(vortex traps surrounded by vortex depletion regions)
postulated by \citet{alp96} and \citet{won01}.
They may be seeded by starquakes,
which create large numbers of fresh lattice dislocations with deep
pinning potentials,
or they may emerge spontaneously in the self-organized critical state,
as successive generations of vortex avalanches traverse the crust.
As the pinned vortices increasingly lag the regular, unpinned array,
a gradient in vortex density is established, and the local Magnus force
on a pinned vortex rises.
When the pinning threshold is overcome, a pinned vortex unpins and
moves abruptly away from the pinning site (stick-slip motion), 
disturbing the local superfluid
velocity field (and hence the Magnus force) appreciably.
Often, this is enough to push neighboring, barely subcritical, pinned vortices 
over their thresholds, triggering an avalanche.
Most vortices in the avalanche rejoin the regular, unpinned array,
and the crust spins up proportionately to compensate.
The time-scale for a vortex to adjust locally to the Magnus and pinning forces
is much shorter than $\nu/\dot{\nu}$, in keeping with (iii).

Classic laboratory experiments on magnetic flux vortices
in a type II superconductor (e.g.\ NbTi) 
immersed in a slowly changing magnetic field,
an exactly analogous system, clearly exhibit properties (iv)--(vi)
\citep{fie95}.
Vortices are expelled mostly in a continuous flow (cf.\ steady spin down)
and occasionally in avalanches (cf.\ glitches).
The distribution of avalanche sizes
is measured to be a power law over several decades,
whose exponent depends on the strength of the applied magnetic field
(which controls the vortex spacing and hence the strength of
the vortex-vortex interaction).
The temporal fluctuation spectrum scales as an inverse power of
frequency at high frequencies.
After initial transients die away, the superconductor fluctuates around a 
self-organised critical state, called the Bean state,
where the Lorentz force acting on each vortex is everywhere equal
to the maximum pinning force.

If the pinning sites are sparsely distributed, 
so that global (mean-field) forces dominate local forces between 
pinned vortex clusters, scale invariance breaks down
\citep{jen98}.
Avalanches still occur, 
but they are distributed narrowly around a characteristic size and lifetime,
involving nearly all the vortices instead of small, independent subsets.
In this regime, avalanches recur quasiperiodically, not stochastically.
Similar behavior is observed when the external driver acts too rapidly,
but this situation never arises in pulsars.

Scale invariant avalanche dynamics and self-organised critical states 
are observed widely elsewhere, in sandpiles
\citep{bak87},
earthquakes
\citep{sor91},
solar flares
\citep{lu91,whe00},
and bursts from soft-gamma-ray
repeaters \citep{gwk00}.
The analogy with pulsar glitches has been pointed out by 
\citet{mor93a} and \citet{car98}
and modeled using a cellular automaton by \citet{mor96}.

\section{Data
 \label{sec:gli3}}
Table \ref{tab:gli1} lists all 285 glitches discovered 
up to the time of writing and known to the authors.
It is compiled from published sources
\citep{she96,lyn00,wan00,hob02,kra03,jan06,mid06,peralta06};
the Australia Telescope National Facility Pulsar Catalogue
\citep{man05},
which can be accessed on-line at 
{\tt http://www.atnf.csiro.au/research/pulsar/psrcat};
and unpublished data communicated privately by 
M.\ Kramer, D.\ Lewis, and A.\ G.\ Lyne.
For each pulsar, the table lists its J2000 coordinates,
and the number of glitches detected
($N_{\rm g}$). The earliest and latest epochs observed
($t_{\rm min}$ and $t_{\rm max}$ respectively) are
recorded separately in Table \ref{tab:gli1b} for the
nine pulsars with $N_{\rm g} > 5$.
An asterisk signifies that segmented data spans are not specified
in the cited references;
in this situation, $t_{\rm min}$ and $t_{\rm max}$ are estimated
by eye from spin-down histories graphed in the cited references,
where available,
or else from the first and last glitches by default.
For each glitch, Table \ref{tab:gli1} lists its epoch,
the fractional increase in spin frequency $\Delta\nu/\nu$,
and one or more bibliographic references.
Uncertainties are quoted as a trailing integer in parentheses,
corresponding to an absolute number of days for $t$
[e.g.\ MJD 51141(248) means MJD $51141\pm 248$]
and an uncertainty in the last signifcant digit for $\Delta\nu/\nu$
[e.g.\ 0.04(2) means $0.04\pm0.02$].
For some newly discovered glitches, the information is incomplete.
Epochs and sizes have been measured for 271 and 250 glitches respectively.
Other parameters, like the healing fraction and
post-glitch relaxation time-scale,
are omitted as they are not analyzed in this paper;
please consult \citet{peralta06} and references therein for a full catalog.

\section{Size distribution
 \label{sec:gli4}}

\subsection{Scale invariance
 \label{sec:gli4a}}
If pulsar glitches are the result of an avalanche process,
their size distribution should be scale invariant in any
individual pulsar, with probability density function
\begin{equation}
 p(\Delta\nu/\nu) \propto (\Delta\nu/\nu)^{-a}~.
 \label{eq:gli1}
\end{equation}
The exponent $a$ is set by the dimensionality
\footnote{The effective dimension need not equal three.
For example, it may equal two in a rectilinear vortex array
or faulting crust,
where the local forces act in the transverse plane,
and three in a turbulent vortex tangle \citep{per05,per06}.}
and symmetries of the local forces,
which are likely to be universal,
and the strength and level of conservation of these forces,
which are functions of temperature and therefore not universal
(see \S\ref{sec:gli2}).
One therefore expects $a$ to differ from pulsar to pulsar.
As a corollary, the aggregate size distribution
drawn from all pulsars
is not expected to be a simple power law of the form (\ref{eq:gli1}).

To test these ideas, we construct the observed cumulative size distributions
of the nine known pulsars with $N_{\rm g} > 5$.
The selection criterion $N_{\rm g}>5$ is arbitrary;
it seeks to limit the impact of random errors while testing as many objects
from Table \ref{tab:gli1} as possible.
We then compare the data against the theoretical cumulative distribution
\begin{equation}
 P(\Delta\nu/\nu)
 =
 \frac{ (\Delta\nu/\nu)^{1-a} - (\Delta\nu/\nu)_{\rm min}^{1-a} }
  { (\Delta\nu/\nu)_{\rm max\phantom{i}}^{1-a} - (\Delta\nu/\nu)_{\rm min}^{1-a} }
 \label{eq:gli2}
\end{equation} 
derived from (\ref{eq:gli1}).
The theoretical distribution is normalized after restricting it
to the domain 
$(\Delta\nu/\nu)_{\rm min} \leq \Delta\nu/\nu 
 \leq (\Delta\nu/\nu)_{\rm max}$,
where 
$(\Delta\nu/\nu)_{\rm min}$ and $(\Delta\nu/\nu)_{\rm max}$ 
are the smallest and largest glitches observed in that pulsar
respectively, quoted in Table \ref{tab:gli1}.
There are more sophisticated ways to choose
$(\Delta\nu/\nu)_{\rm min}$ and $(\Delta\nu/\nu)_{\rm max}$,
which we consider further below, but this is a conservative starting point.

\begin{figure}
\includegraphics[scale=0.9]{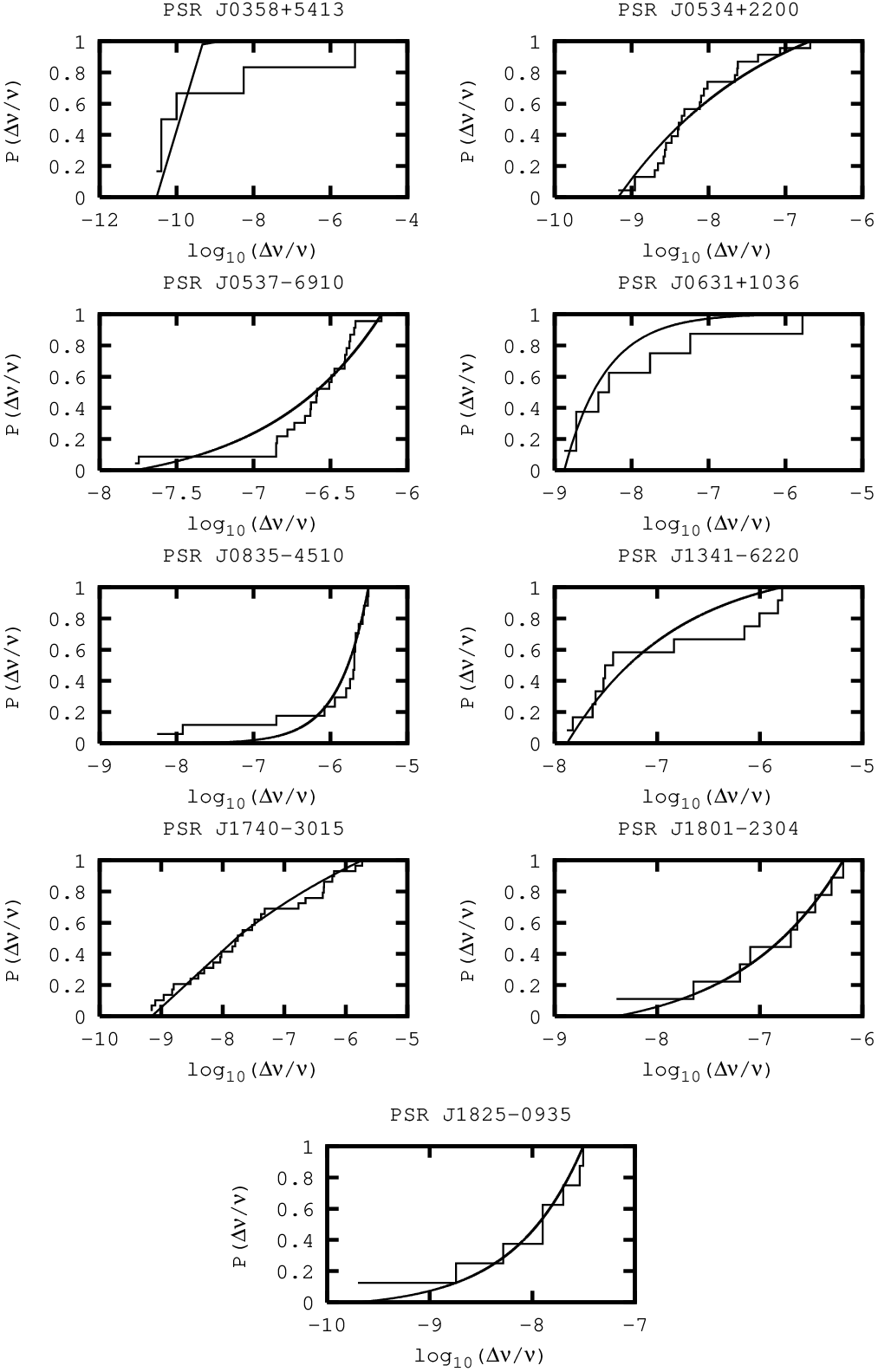}
\caption{Cumulative distribution of fractional glitch sizes $\Delta\nu/\nu$ in
the nine pulsars that have glitched more than five times. The observational data ({\em histogram})
are plotted together with the best power-law fit ({\em solid curve}) given by (\ref{eq:gli2}), with
$(\Delta\nu/\nu)_{\rm min}$ and $(\Delta\nu/\nu)_{\rm max}$ taken from Table \ref{tab:gli1}
and $a$ taken from Table \ref{tab:gli2}.}
\label{fig:gli1}
\end{figure}

For each object, we choose $a$ to minimize the Kolmogorov-Smirnov (K-S)
statistic $D$, i.e.\ the maximum unsigned distance between the curves.
The numerical results are recorded in Table \ref{tab:gli2},
while the measured and theoretical cumulative distributions are plotted
together in Figure \ref{fig:gli1}.
(Cumulative distributions are free of binning bias.)
The goodness of the fit at the optimal value of $a$ is characterized
by $P_{\rm KS}$, defined such that $1-P_{\rm KS}$ equals
the probability that the K-S null hypothesis
(that the two data sets are drawn from the same underlying distribution) 
is false.
\footnote{The K-S test is most sensitive to discrepancies near the median.
An alternative test, based on Kuiper's statistic, mitigates this bias
\citep{pre86}. 
It will be worth implementing when more data become available.
The K-S test is also inefficient if the underlying probability density function
contains a narrow notch, where the probability vanishes.
Again, there is insufficient data at hand to look for such a notch;
it is difficult to find in a cumulative distribution, 
and the probability density function is biased by binning when $N_{\rm g}$ is small.}
The 1-$\sigma$ lower and upper bounds $a_-$ and $a_+$ mark the 
range of $a$ where the null hypothesis is rejected with less than
$68\%$ confidence. Note that the interval $[a_-,a_+]$ is 
asymmetric about the optimal $a$ and widest for the best fits.

The results in Table \ref{tab:gli2} confirm what is apparent by eye
from Figure \ref{fig:gli1}: the null hypothesis that
the size distribution is described by a power law for all nine pulsars
with $N_{\rm g} > 5$ is not ruled out at the 1-$\sigma$ level of confidence.
In turn, this is consistent with the avalanche hypothesis.
However, in two objects, namely PSR J0537$-$6910 and PSR J0835$-$4510,
the agreement is marginal.
Interestingly, these two objects are also the only ones discovered so far
that are believed to glitch quasiperiodically
\citep{lyn96,mid06}.

\subsection{PSR J0537$-$6910 and PSR J0835$-$4510
 \label{sec:gli4b}}
Quasiperiodicity is a
natural feature of avalanche dynamics when mean-field forces 
overwhelm local interactions, as described in \S\ref{sec:gli2}.
We explore its manifestation in glitch waiting times in \S\ref{sec:gli5}.
With respect to glitch sizes, we note that avalanches in the
quasiperiodic regime tend to be distributed narrowly around a 
characteristic size
\citep{jen98}.
This can be modeled crudely by adding a term proportional to 
$\delta[\Delta\nu/\nu - (\Delta\nu/\nu)_{\rm c}]$
to (\ref{eq:gli1}), viz.\
\begin{equation}
 p(\Delta\nu/\nu)
 =
 C_{\rm s} (\Delta\nu/\nu)^{-a} + 
 C_{\rm p} \delta[\Delta\nu/\nu - (\Delta\nu/\nu)_{\rm c}]~,
 \label{eq:gli3a}
\end{equation}
where $(\Delta\nu/\nu)_{\rm c}$ denotes the characteristic size,
and the scale invariant and quasiperiodic components are weighted by
the constants $C_{\rm s}$ and $C_{\rm p}$ respectively.
Normalization fixes $C_{\rm s}$ in terms of $C_{\rm p}$
(or vice versa), with $C_{\rm s} + C_{\rm p} \neq 1$ in general.
The associated cumulative size distribution is given by
\begin{eqnarray}
 P(\Delta\nu/\nu)
 & = &
 C_{\rm p} H[ \Delta\nu/\nu - (\Delta\nu/\nu)_{\rm c} ]
 \nonumber \\
 & &
 + 
 \frac{ (1-C_{\rm p}) [ (\Delta\nu/\nu)^{1-a} - (\Delta\nu/\nu)_{\rm min}^{1-a} ] }
  { (\Delta\nu/\nu)_{\rm max\phantom{i}}^{1-a} - (\Delta\nu/\nu)_{\rm min}^{1-a} }~,
 \label{eq:gli3b}
\end{eqnarray}
where $H(\cdot)$ denotes the Heaviside step function.

\begin{figure}
\plottwo{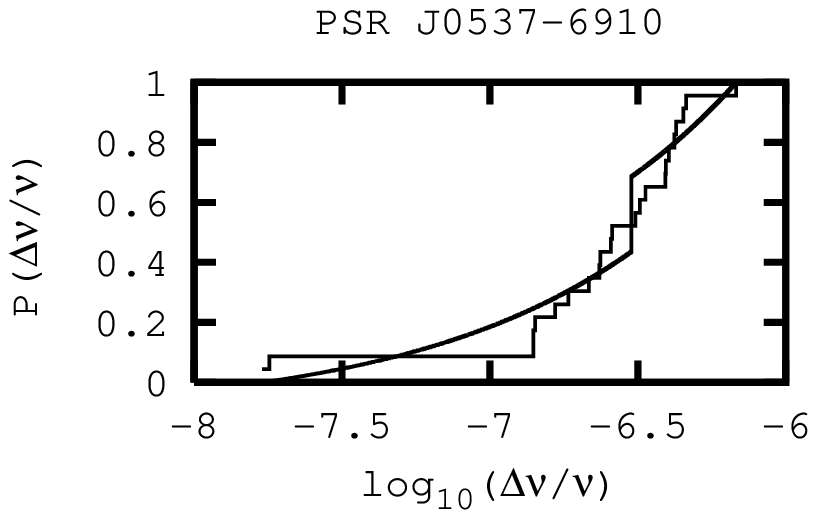}{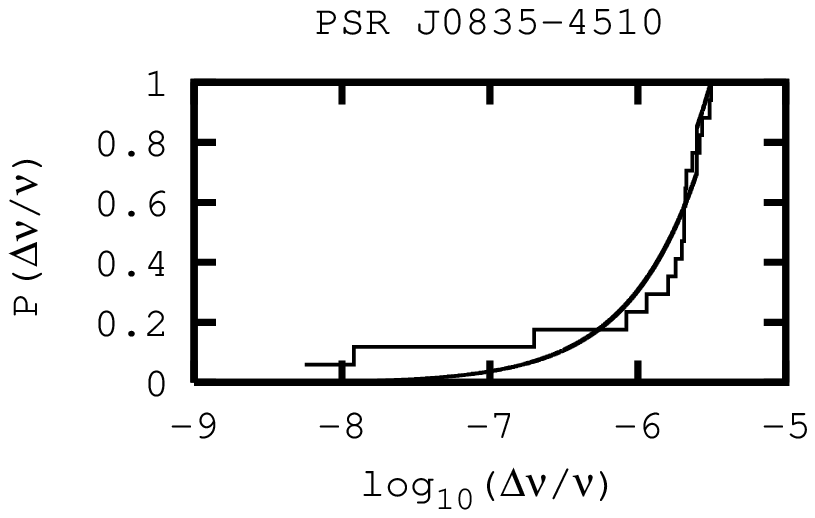}
\caption{Cumulative distribution of fractional glitch sizes $\Delta\nu/\nu$ for 
the two pulsars which have a quasiperiodic component.
The observational data ({\em histogram}) are plotted together with the best two-component fit ({\em solid curve}) given by (\ref{eq:gli3b}), with
$(\Delta\nu/\nu)_{\rm min}$ and $(\Delta\nu/\nu)_{\rm max}$ taken from Table \ref{tab:gli1},
and $a$, $C_p$ and $(\Delta\nu/\nu)_c$ taken from Table \ref{tab:gli3}.
}
\label{fig:gli2}
\end{figure}

Parameters determined by fitting (\ref{eq:gli3b}) to the data are recorded
in Table \ref{tab:gli3},
while the corresponding measured and theoretical cumulative distributions
are plotted together in Figure \ref{fig:gli2}.
The fits are much improved,
with $C_{\rm p}\approx 0.2$ in both objects
--- although, to be fair, the delta-distributed component is not 
strictly required, at least not 
at the 1-$\sigma$ level.
Importantly, the delta-distributed component contains
only $\sim20\%$ of the glitches, not all of them.
This is consistent with the historical interpretation of the pulsar data
\citep{lyn95,mar04}.
It is also seen in self-organized critical systems like sandpiles, 
where large, system-spanning, quasiperiodic avalanches
of a characteristic size are interspersed with small, randomly timed
avalanches, which are power-law distributed
\citep{ros93,jen98}.

\subsection{Upper and lower cut-offs
 \label{sec:gli4c}}
Strictly speaking, it is incorrect to normalize $P(\Delta\nu/\nu)$
by choosing 
$(\Delta\nu/\nu)_{\rm min}$ and $(\Delta\nu/\nu)_{\rm max}$ 
to be the smallest and largest glitches observed in a pulsar
respectively.
A better choice of $(\Delta\nu/\nu)_{\rm min}$ is the actual
resolution of the timing experiment, 
which varies with object and epoch. 
\citet{jan06} simulated detection of a microglitch in a noisy time series
and obtained $(\Delta\nu/\nu)_{\rm min}=1\times 10^{-11}$.
Usually, this information is not provided explicitly 
and must be estimated from the size uncertainties quoted for
detected glitches.
All the same, the smallest glitch observed is likely to be a
reasonable estimate of $\Delta\nu_{\rm min}$,
because the occurrence probability increases steeply as $\Delta\nu$ decreases, 
according to Table \ref{tab:gli2}.
On the other hand, $(\Delta\nu/\nu)_{\rm max}$ is limited by the
total observing time.
Its true value exceeds the largest glitch observed, but not by much, 
because the occurrence probability decreases steeply as $\Delta\nu$ increases.

To quantify these effects, we allow 
$(\Delta\nu/\nu)_{\rm min}$ to vary between 0.5 and 1.0 times the
smallest glitch size observed,
$(\Delta\nu/\nu)_{\rm max}$ to vary between 1.0 and 2.0 times the
largest glitch size observed,
and fit equation (\ref{eq:gli3b}) again to the data.
For every object,
$(\Delta\nu/\nu)_{\rm min}$ and $(\Delta\nu/\nu)_{\rm max}$
shift only slightly,
and $a$ stays within the range $[a_-,a_+]$ in Tables \ref{tab:gli2} and \ref{tab:gli3}.
This confirms that the smallest and largest glitches
provide reasonable estimates of
$(\Delta\nu/\nu)_{\rm min}$ and $(\Delta\nu/\nu)_{\rm max}$.
At the 1-$\sigma$ level, the constrained and unconstrained fits
are both consistent with the data. 

\subsection{Aggregate distribution
 \label{sec:gli4d}}
\begin{figure}
\plottwo{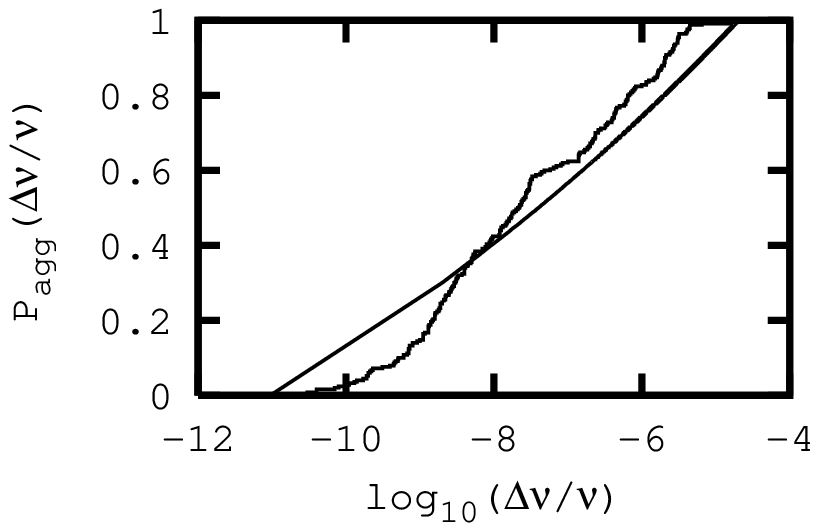}{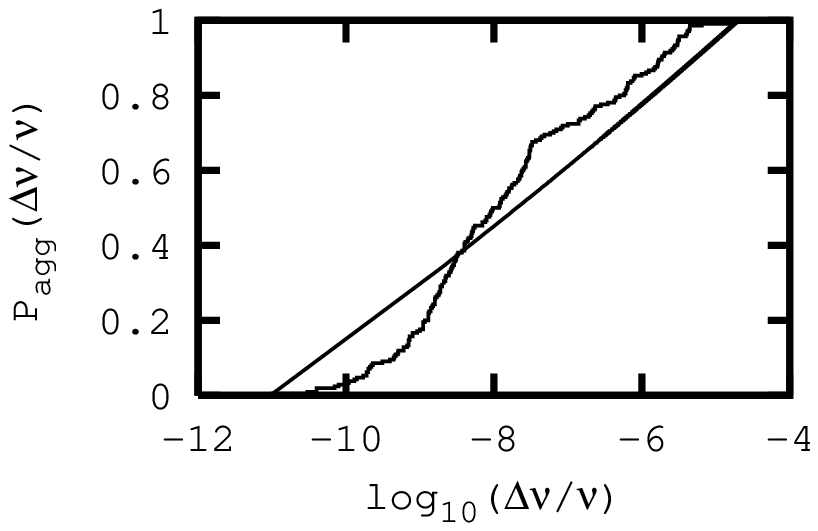}
\caption{Aggregate cumulative distribution of fractional glitch sizes $\Delta\nu/\nu$ 
for all glitching pulsars ({\em left panel}) and for all glitching pulsars
except PSR J0537$-$6910 and PSR J0835$-$4510, which have a quasiperiodic
component ({\em right panel}). The observational data ({\em histogram}) 
are plotted together with the best power-law fit ({\em solid curve})
given by (\ref{eq:gli2}), for the best fit parameters in \S\ref{sec:gli4d}.}
\label{fig:gli3}
\end{figure}

Figure \ref{fig:gli3} displays the cumulative size distribution for
the glitch population in aggregate,
together with the best power-law fit of the form (\ref{eq:gli2}).
The fit is poor.
When all 250 glitches with measured sizes are included, 
the best fit corresponds to
$a=0.96$, $(\Delta\nu/\nu)_{\rm min}=9.5\times 10^{-12}$,
$(\Delta\nu/\nu)_{\rm max}=2.0\times 10^{-5}$, 
and $P_{\rm KS}=7.1\times 10^{-4}$.
When the glitches from PSR J0537$-$6910 and PSR J0835$-$4510 are excluded,
the best fit corresponds to
$a=0.98$ and $P_{\rm KS}=3.2\times 10^{-4}$,
with $(\Delta\nu/\nu)_{\rm min}$ and $(\Delta\nu/\nu)_{\rm max}$
as before. 
Either way, we can state confidently that
the theoretical and observed data are drawn from different
underlying distributions.
This is not surprising; the results in Figure \ref{fig:gli1}
and Table \ref{tab:gli2} demonstrate clearly that the size distribution
in individual pulsars is consistent with being scale invariant, but that $a$
differs from object to object.
Hence the aggregate distribution is expected to be a weighted sum of 
power laws, not a pure power law.
Accordingly, the size distribution in individual pulsars
is a more direct probe of glitch physics than the aggregate distribution
\citep{lyn00}.
The aggregate distribution can be inverted, in principle,
to determine how $a$ is distributed across the pulsar population.
We defer this exercise until better historical estimates 
of $(\Delta\nu/\nu)_{\rm min}$ and $(\Delta\nu/\nu)_{\rm max}$, 
and more data, become available.

\citet{jan06} claimed that the glitches in PSR J1740$-$3015 are
drawn from a flat size distribution in $\log (\Delta\nu/\nu)$,
i.e. $a=1$,
with $P_{\rm KS}=0.902$.
This agrees with the results in Table \ref{tab:gli2}.

\citet{lyn00} noted some evidence for an excess of large glitches,
which is corroborated to some extent by Figure \ref{fig:gli3}a.
However, the excess largely disappears when the quasiperiodic
glitchers are excluded, as in Figure \ref{fig:gli3}b.
Large glitches do not originate preferentially from any
particular class of object.
While it is true that the most active objects
(e.g.\ PSR J0537$-$6910 and PSR J0835$-$4510)
experience relatively large and narrowly distributed glitches,
with $\Delta\nu/\nu > 10^{-7}$,
other active objects (e.g. PSR J1740$-$3015) experience a mix of small
and large events,
and there are several objects (e.g.\ PSR J1806$-$4212) which have only
glitched once, with $\Delta\nu/\nu > 10^{-5}$ for that single glitch.
Furthermore, although PSR J0534$+$2200 is sometimes portrayed as unusual
for not experiencing large glitches, its size distribution is actually
relatively flat ($a\approx 1.2$).
There is every reason to expect that it will experience large glitches 
in the future, but these events will be slow in coming,
because PSR J0534$+$2200 builds up 
differential rotation between the crust and superfluid
at a relatively slow rate,
as we show in \S\ref{sec:gli5}. 
\footnote{
\citet{won01} argued that the relative angular acceleration of the crust
and superfluid in the Crab, inferred from the activity parameter,
is much smaller than expected given the large 
$\Delta\dot{\nu}/\dot{\nu}\sim 10^{-4}$
observed during glitches.
This paradox is resolved if most of the differential rotation
is being stored temporarily, in advance of a large glitch in the future.
}

\section{Waiting-time distribution
 \label{sec:gli5}}

\subsection{Poisson process
 \label{sec:gli5a}}
If pulsar glitches are the result of an avalanche process,
they should be statistically independent events.
To understand why, recall that a system in a self-organized 
critical state configures itself into many metastable stress reservoirs
insulated by relaxed zones (\S\ref{sec:gli2}).
Every avalanche relaxes one reservoir,
typically occupying a small fraction of the system, 
and the next avalanche occurs at random, typically far from its predecessor.
There is essentially no interference between successive avalanches;
this is verified empirically in tests with cellular automata
\citep{jen98}.
Avalanches in the tail of the size distribution, which relax the whole system,
are an (extremely rare) exception.

Given statistical independence, and assuming that the system is driven 
at a constant (mean) rate, the avalanche model predicts that the
time between successive glitches, $\Delta t$, termed the waiting time,
obeys Poisson statistics
\citep{jen98,whe00}.
Hence, in any individual pulsar, the waiting-time distribution is exponential,
with probability density function
\begin{equation}
 p(\lambda, \Delta t)
 =
 \lambda \exp(-\lambda \Delta t)~.
\label{eq:gli4}
\end{equation}
The mean glitching rate $\lambda$ is different for every pulsar. 
It depends on the rate at which differential rotation builds up
between the superfluid and the crist ($\propto \dot{\nu}$)
as well as the capacity to store the differential rotation
(e.g.\ strength of pinning, rate of vortex creep, shear modulus
of the crust).
The storage capacity is presumably controlled by thermodynamic variables
like temperature, as well as the inhomogeneous nuclear structure of the crust.
We do not expect $\lambda$ to change appreciably during four decades
of pulsar timing. In principle, however, as more data are collected in future,
this claim can be tested by using a Bayesian blocks algorithm
to divide the time series into a sequence of Poisson processes with
piecewise-constant rates
\citep{sca98,whe00,con03}.
\footnote{
It is tempting to assume that $\lambda$ is constant over decades,
because the thermodynamic variables that control
storage capacity (e.g.\ temperature) are nearly constant on
such a time-scale. 
Yet the Sun provides a cautionary counterexample:
the dynamics of subphotospheric turbulence, and hence the rate of
solar flaring, vary with the 11-yr solar cycle \citep{whe00}.
}

The avalanche model makes a further powerful prediction.
Suppose the system tends to a stationary, self-organised critical state,
in which global driving balances local release in a time-averaged sense
[property (vi), \S2],
i.e.\ there is no secular accumulation or leakage of stress.
Stationarity implies that the mean waiting time 
$\langle \Delta t \rangle = \lambda^{-1}$, 
multiplied by the rate at which crust-superfluid differential rotation 
builds up ($\epsilon\dot{\nu}$),
equals the mean glitch size $\langle \Delta\nu \rangle$,
i.e.\ 
\begin{equation}
 \lambda = - \epsilon\dot{\nu} / \langle \Delta\nu \rangle~.
 \label{eq:gli5}
\end{equation}
Here, $2\pi\epsilon\dot{\nu}$ is the relative angular acceleration
of the crust and superfluid. 
Importantly, glitch data allow $\epsilon$ to be measured directly
in principle \citep{won01}. However, there is a serious
question as to whether stationarity is achieved
in practice, during the $40$ yr that a typical
pulsar has been observed. We discuss this issue
further below.

To test the above ideas, we compare the 
observed cumulative waiting-time distributions
of the same nine pulsars as in \S\ref{sec:gli4}, with $N_{\rm g} > 5$,
against the theoretical cumulative distribution.
In order to make the comparison fairly, 
we must first adjust for observational selection effects.
Any given obesrvation can detect waiting times in a range
$\Delta t_{\rm min} \leq \Delta t \leq \Delta t_{\rm max}$.
The upper limit $\Delta t_{\rm max}$ is set by the total data span
available for that pulsar,
i.e.\ $\Delta t_{\rm max} = t_{\rm max} - t_{\rm min}$.
The lower limit $\Delta t_{\rm min}$ is different at different epochs.
It is set by the gap between data spans in which a glitch
is localized.
For small glitches, the glitch epoch is determined by requiring
continuity of pulse phase across the glitch.
For larger glitches, where the phase winding number is ambiguous,
the epoch is taken to be halfway between the bounding observations
\citep{wan00}.
Either way, $\Delta t_{\rm min}$ is different for each glitch,
and is twice the absolute value of the epochal uncertainty 
quoted in Table \ref{tab:gli1}
\citep{lyn00,wan00,jan06,mid06}.
Let $f(\Delta t_{\rm min})d(\Delta t_{\rm min})$ 
be the observing-time-weighted probability that, when a glitch occurs,
$\Delta t_{\rm min}$ lies in the range
$[ \Delta t_{\rm min},
 \Delta t_{\rm min} + d(\Delta t_{\rm min}) ]$,
and let the smallest and largest values of $\Delta t_{\rm min}$
be $\Delta t_{\rm min}^{(<)}$ and $\Delta t_{\rm min}^{(>)}$ respectively,
recorded in Table \ref{tab:gli4} for the nine pulsars
with $N_{\rm g} > 5$.
Then the cumulative waiting-time distribution is given by
\begin{eqnarray}
 P(\lambda, \Delta t) 
 & = &
 \int_{\Delta t_{\rm min}^{(<)}}^{\Delta t_{\rm min}^{(>)}}
 d(\Delta t_{\rm min}') \, f(\Delta t_{\rm min}') \nonumber \\
& & \times \int_{\Delta t_{\rm min}'}^{\Delta t_{\rm max}} 
 d(\Delta t') \, p(\lambda, \Delta t')~,
 \label{eq:gli6}
 \\
 & = &
 \frac{1}{N_{\rm g}}
 \sum_{\Delta t_{\rm min}=\Delta t_{\rm min}^{(<)}}^{\Delta t_{\rm min}^{(>)}}
 \frac{ \exp(-\lambda\Delta t_{\rm min}) - \exp(-\lambda\Delta t) }
  { \exp(-\lambda\Delta t_{\rm min}) - \exp(-\lambda\Delta t_{\rm max}) }~,
\nonumber \\
 \label{eq:gli7}
& & 
\end{eqnarray}
where each glitch is weighted equally in the sum in (\ref{eq:gli7})
as a first approximation. 

\begin{figure}
\includegraphics[scale=0.9]{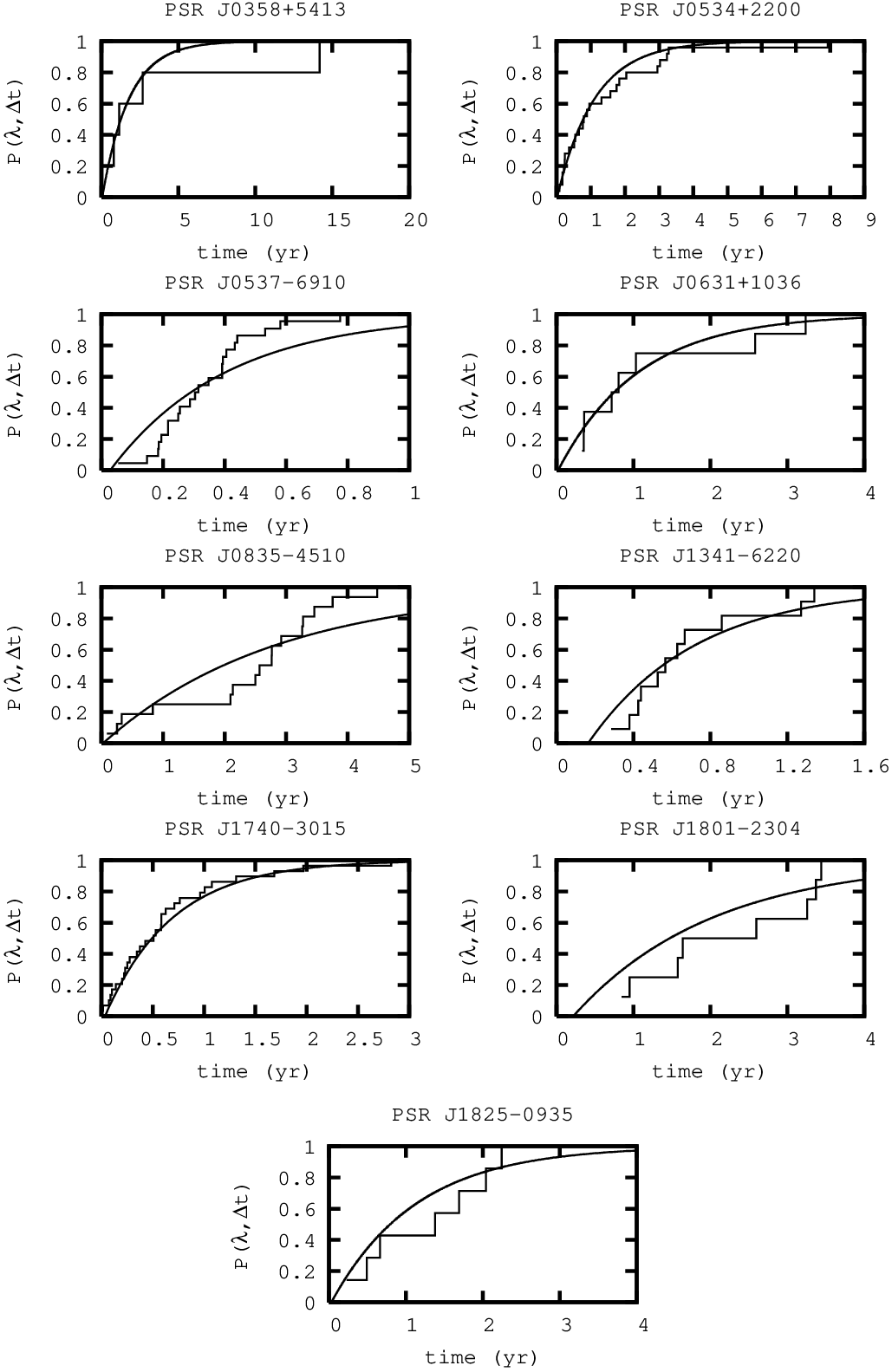}
\caption{
Cumulative distribution of 
glitch waiting times $\Delta t$ (measured in {\rm yr}) for
the nine pulsars that have glitched more than five times.
The observational data ({\em histogram}) are plotted together with the best Poisson fit ({\em solid curve}) given by 
(\ref{eq:gli7}), with
$\Delta t_{\rm min}$ taken from Table \ref{tab:gli1} (twice
the epochal uncertainty), $\Delta t_{\rm max}$ from Table \ref{tab:gli1b}, and
$\lambda$ from Table \ref{tab:gli5}.}
\label{fig:gli4}
\end{figure}

In Figure \ref{fig:gli4}, we plot as cumulative histograms the measured
waiting-time distributions of the nine pulsars in Figure \ref{fig:gli1}.
The theoretical curves (\ref{eq:gli7}) are overlaid,
with $\Delta t_{\rm min}$ and $\Delta t_{\rm max}$ 
chosen according to the second through fourth columns in Table \ref{tab:gli4}.
For each object, we choose $\lambda$ to minimize the K-S statistic $D$.
The fitting parameters are displayed in the fifth through eighth columns
in Table \ref{tab:gli4}.
As in \S\ref{sec:gli4},
the goodness of the fit is characterized by the K-S probability $P_{\rm KS}$,
with $P_{\rm KS} < 0.32$ in the interval $[\lambda_-,\lambda_+]$.

For all nine pulsars in Figure \ref{fig:gli4} and Table \ref{tab:gli4},
the null hypothesis that
the waiting-time distribution is described by Poisson statistics
is not ruled out at the 1-$\sigma$ level.
The data are therefore consistent with an avalanche process.
An exponential waiting-time distribution was first postulated by
\citet{won01}
for PSR J0534$+$2200, 
based on timing data up to and including the glitch on MJD 51452.
These authors obtained 
$\lambda = 0.53\,{\rm yr^{-1}}$ and $P_{\rm KS}=0.7$,
marginally outside the 1-$\sigma$ range in Table \ref{tab:gli4}.
The data analyzed here confirm that waiting times are
consistent with Poisson statistics
in several glitching pulsars,
affording a key insight into the physics of the glitch mechanism. 
The implications of this result are discussed further in \S\ref{sec:gli6}.

\subsection{Quasiperiodicity
 \label{sec:gli5b}}
For seven pulsars in Figure \ref{fig:gli4}, 
the Poisson distribution affords an excellent fit, 
both formally and by eye.
However, for PSR J0537$-$6910 and PSR J0835$-$4510,
the fits are marginal at the 1-$\sigma$ level.
[Indeed, in an earlier analysis of PSR J0835$-$4510,
\citet{won01} excluded a Poisson distribution with $96\%$ confidence,
on the basis of fewer data.]
These are the same two objects whose size distributions are exceptional,
and which are observed to glitch quasiperiodically.

Taking the same approach as in \S\ref{sec:gli4}, 
we model the quasiperiodicity crudely by adding a periodic
component to (\ref{eq:gli4}), viz.
\begin{equation}
 p(\lambda,\Delta t) 
 =
 C_{\rm s}' \lambda \exp(-\lambda\Delta t)
 + C_{\rm p}' \delta ( \Delta t - \Delta t_{\rm c} )~.
 \label{eq:gli8}
\end{equation}
In (\ref{eq:gli8}), $C_{\rm s}'$ and $C_{\rm p}'$ are the 
normalized relative weights
of the Poisson and periodic components respectively,
and $\Delta t_{\rm c}$ is the characteristic period.
The associated cumulative distribution, weighted by $\Delta t_{\rm min}$,
is obtained by substituting (\ref{eq:gli8}) into (\ref{eq:gli6}),
yielding
\begin{eqnarray}
 P(\lambda,\Delta t)
 & = &
 \frac{1}{N_{\rm g}}
 \sum_{\Delta t_{\rm min}=\Delta t_{\rm min}^{(<)}}^{\Delta t_{\rm min}^{(>)}}
 \left\{ 
  C_{\rm p}' H(\Delta t - \Delta t_{\rm c})
  \phantom{\frac{1}{2}}
 \right.
 \nonumber \\
 & & 
 \left.
 + 
 \frac{ (1-C_{\rm p}') [ \exp(-\lambda\Delta t_{\rm min}) - \exp(-\lambda\Delta t) ] }
   { \exp(-\lambda\Delta t_{\rm min}) - \exp(-\lambda\Delta t_{\rm max}) }
 \right\}~. \nonumber \\
& & 
 \label{eq:gli9}
\end{eqnarray}

\begin{figure}
\plottwo{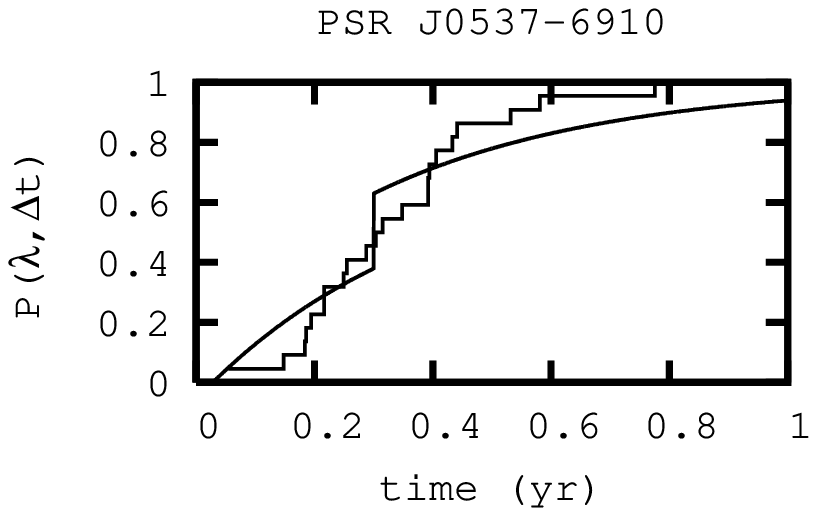}{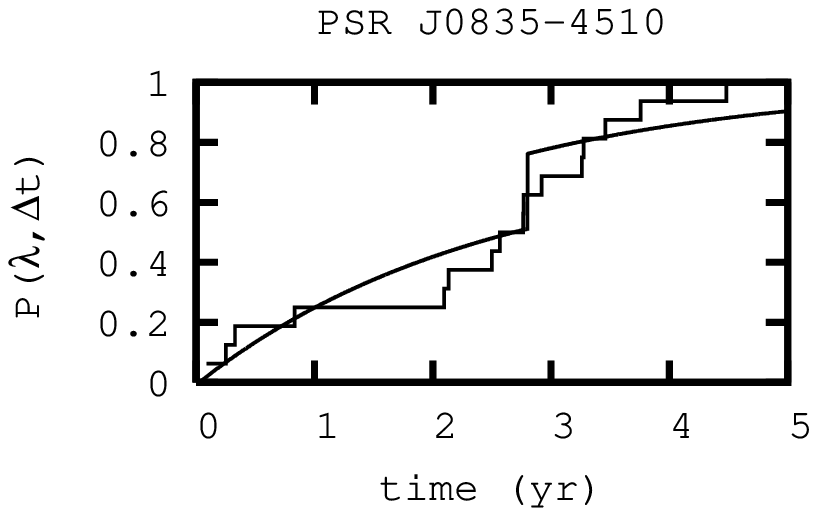}
\caption{
Cumulative distribution of
glitch waiting times $\Delta t$ (measured in {\rm yr}) for the two pulsars which have a quasiperiodic component.
The observational data ({\em histogram}) are plotted together with the best two-component fit ({\em solid curve}), given 
by (\ref{eq:gli9}), with $\Delta t_{\rm min}$ taken from Table \ref{tab:gli1}
(twice the epochal uncertainty),  $\Delta t_{\rm max}$ from Table \ref{tab:gli1b},
and $\lambda$, $C_p'$, and $\Delta t_c$ taken from Table \ref{tab:gli1}.  }
\label{fig:gli5}
\end{figure}

The two-component model (\ref{eq:gli8}) yields improved fits 
to the data, with $C_{\rm p}' \approx 0.25$ for both objects.
The fits are graphed with the data in Figure \ref{fig:gli5},
and the best-fit parameters are recorded in Table \ref{tab:gli5}.
We obtain $\Delta t_{\rm c} = (0.3\pm0.1)\,{\rm yr}$ 
and $\Delta t_{\rm c} = (2.8\pm0.1)\,{\rm yr}$ 
for PSR J0537$-$6910 and PSR J0835$-$4510 respectively,
in accord with previous authors
\citep{lyn96,mid06}.
Significantly, the data imply $C_{\rm p} \approx C_{\rm p}'$.
In other words,
the delta-distributed component accounts for the same fraction
of the size and waiting-time distributions, even though the sizes
and waiting times are statistically independent.
This raises confidence in the model and suggests that a
quasiperiodic component is indeed present and distinct.
It also suggests that the quasiperiodic component coexists with
the Poisson component, instead of completely displacing it.
Vela, for example, is likely to possess an extensively connected network
of capacitive elements, but it also contains smaller subnetworks
that are disconnected from the main network;
cf.\ \citet{alp96}.
This is natural for an avalanche process,
as noted in \S\ref{sec:gli4b}
\citep{ros93,jen98}.

\subsection{Mean rate
 \label{sec:gli5c}}
Stationarity of the avalanche model over long time intervals
implies a relation between
the Poisson rate, driving rate, and mean glitch size,
given by (\ref{eq:gli5}).
Unfortunately, for $a < 2$,
$\langle \Delta \nu \rangle$ is dominated by large glitches
near the upper cut-off of $P(\Delta \nu/\nu)$:
\footnote{For $1 < a < 2$, $\langle \Delta \nu \rangle$
is dominated by the upper cut-off, but the normalization
of $p(\Delta \nu/\nu)$ is dominated by the lower cut-off.}
\begin{equation}
\label{eq:new_equation}
\langle \Delta \nu \rangle  = \left|\frac{a-1}{a-2}\right| \left\{ \begin{array}{ll}
       \displaystyle  \Delta \nu_{\rm upper} & a > 2 \\
        \displaystyle  \left(\frac{\Delta \nu_{\rm upper}}{\Delta \nu_{\rm lower}}\right)^{a-1} \Delta \nu_{\rm upper} & 1 < a < 2 \\
\displaystyle \Delta \nu_{\rm upper} & a < 1.
     \end{array}\right.
\end{equation}
In (\ref{eq:new_equation}), $\Delta \nu_{\rm lower}$
and $\Delta \nu_{\rm upper}$ are the physical lower and upper 
cut-offs of the probability distribution function (\ref{eq:gli1}).
As large glitches are rare, stationarity is not achieved during
the $40$-yr interval over which a typical pulsar is observed; the
largest observed size, $(\Delta \nu/\nu)_{\rm max}$, cannot be equated
reliably with the maximum size allowed physically.
Likewise, $\langle \Delta \nu\rangle$ is approximated poorly by the
mean of the observed glitches. In practice, therefore, it is impossible
to estimate $\langle \Delta \nu\rangle$ without much longer monitoring.

Physically, $\epsilon\dot{\nu}$ is the rate at which
differential rotation builds up between the crust and superfluid.
Hence, in the vortex unpinning model, $\epsilon$ gives the time-averaged
fraction of pinned vortices or capacitive elements.
We can use equation (\ref{eq:new_equation}) to place limits
on $\epsilon$, at least in principle. \footnote{The error in
$\Delta \nu$ is $d \langle \Delta \nu \rangle /da$ multiplied
by the error in $a$.}
For example, the inequalities $\Delta \nu_{\rm lower} < \Delta \nu_{\rm min}$
and $\Delta \nu_{\rm max} < \Delta \nu_{\rm upper} < \nu$
must always be satisfied. For PSR J0358$+$5413, assuming
$a=2.4$, we find $\langle \Delta \nu\rangle/\nu \leq 1.1 \times 10^{-10}$
and hence $\epsilon \leq 7 \times 10^{-5}$. This is lower
than previous estimates of the pinned fraction for objects
of that age, but in line with previous estimates of the pinned
fraction in young objects like the Crab \citep{lyn00,won01}.
For the remaining eight objects, the above inequalities lead
to upper limits on $\epsilon$ which are greater
than unity and hence not useful. As a crude experiment,
we check the result of setting $\Delta \nu_{\rm upper} = 2 \times 10^{-4}$,
the largest glitch observed in any pulsar over the last $40$ years,
for every pulsar.
We obtain five slightly more useful upper limits
($\epsilon \leq 4 \times 10^{-2}$, $0.2$, $0.8$, $0.8$, and $0.7$ for
PSR J0534$+$2200, PSR J0631$+$1036, PSR J0835$-$4510, PSR J1341$-$6220, and
PSR J1740$-$3015 respectively). However, we emphasize that these values
are still problematic, because there is no guarantee that a total effective
observation interval of $40$ yr $\times$ $101$ pulsars 
is long enough in aggregate for
stationarity to be observed. Moreover, these values are based
on assuming that all pulsars have the same physical $\Delta \nu_{\rm upper}$,
which is not necessarily the case.

Figure \ref{fig:gli6} displays the cumulative histogram of 
Poisson rates derived from the best-fit waiting-time distributions 
in Figures \ref{fig:gli4} and \ref{fig:gli5}.
Let $q(\lambda)$ denote the rate probability density function,
such that $q(\lambda)d\lambda$ is the probability that the mean rate
lies in the interval $[\lambda,\lambda+d\lambda]$ in a given pulsar.
There is no obvious theoretical reason to prefer a particular 
analytic form of $q(\lambda)$,
which is controlled by the physics of the global driver,
not the scale invariant avalanche dynamics.
In addition, the data in Figure \ref{fig:gli6} are insufficient to 
specify the analytic form of $q(\lambda)$ uniquely.
However, motivated by the rate distribution observed in solar flares,
which is measured reliably to be exponential
\citep{whe00},
we find that a distribution of the form
\begin{equation}
 q(\lambda) 
 =
 \langle \lambda \rangle^{-1}
 \exp(-\lambda / \langle \lambda \rangle )
 \label{eq:gli10}
\end{equation}
fits the data satisfactorily, with
$\langle \lambda \rangle = 1.3_{-0.6}^{+0.7}\,{\rm yr^{-1}}$
including quasiperiodic glitchers ({\em left panel})
or $\langle \lambda \rangle = 1.2_{-0.4}^{+0.5}\,{\rm yr^{-1}}$
excluding quasiperiodic glitchers ({\em right panel}).
Formally, the K-S probabilities are $P_{\rm KS} = 0.9946$
and $0.82$ respectively.

The distribution is incompletely sampled below an effective
minimum rate $\lambda_{\rm min}$,
which is set by $\Delta t_{\rm max}$.
To illustrate, if we proclaim five glitches (arbitrarily) to be 
the minimum number needed for a reliable determination of $\lambda$,
we obtain 
$\lambda_{\rm min} = 5 \Delta t_{\rm max}^{-1} 
 \sim 0.2\,{\rm yr^{-1}}$.
Careful modeling of this observational bias is deferred
to a future paper. We describe a first attempt in \S\ref{sec:gli5d}.

\subsection{Aggregate distribution
 \label{sec:gli5d}}
The nine pulsars in Figure \ref{fig:gli4} account for only 108 out of a 
total of 285 observed glitches.
Most glitching pulsars have only glitched once or twice,
but they still contribute statistical information regarding 
waiting times, the former via lower limits on $\Delta t$.
While these data cannot usefully constrain $P(\lambda,\Delta t)$
in individual pulsars, they feed into the aggregate waiting-time distribution
and hence constrain $q(\lambda)$ more tightly than in \S\ref{sec:gli5c}.

\begin{figure}
\plottwo{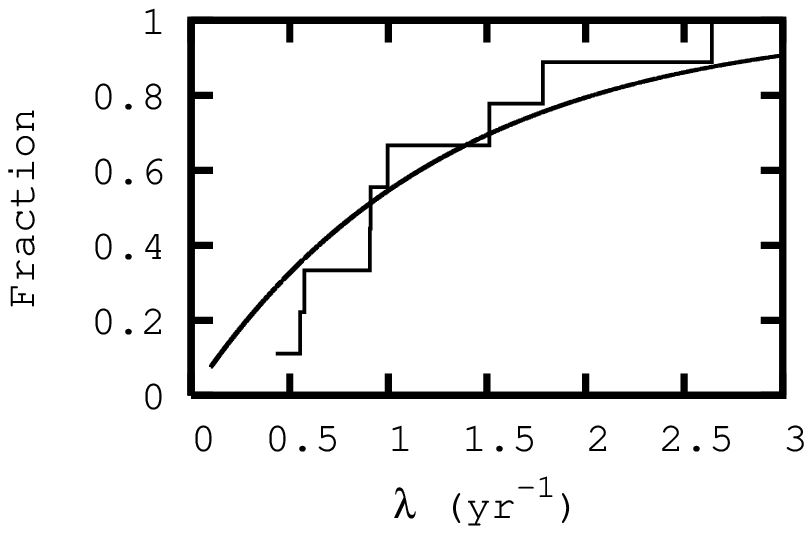}{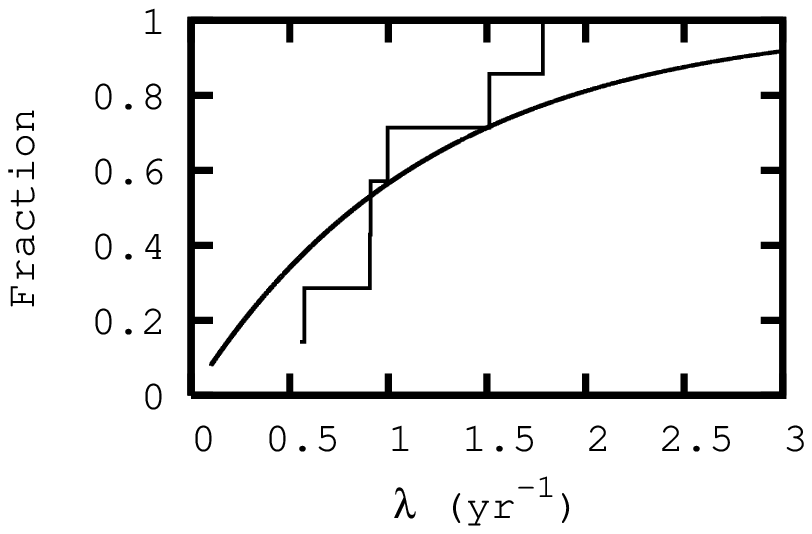}
\caption{Cumulative distribution of mean glitching rate $\lambda$ (measured in {\rm yr}$^{-1}$)
for the nine pulsars that have glitched more than five times
({\em left panel}) and excluding the two quasiperiodic glitchers ({\em right panel}),
showing the observational data ({\em histogram}) and the best fit obtained from (\ref{eq:gli10}) ({\em solid curve}),
with $\langle \lambda \rangle = 1.3_{-0.6}^{+0.7}\,{\rm yr^{-1}}$ ({\em left panel}) and 
$\langle \lambda \rangle = 1.2_{-0.4}^{+0.5}\,{\rm yr^{-1}}$ ({\em right panel}).}
\label{fig:gli6}
\end{figure}

\begin{figure}
\plottwo{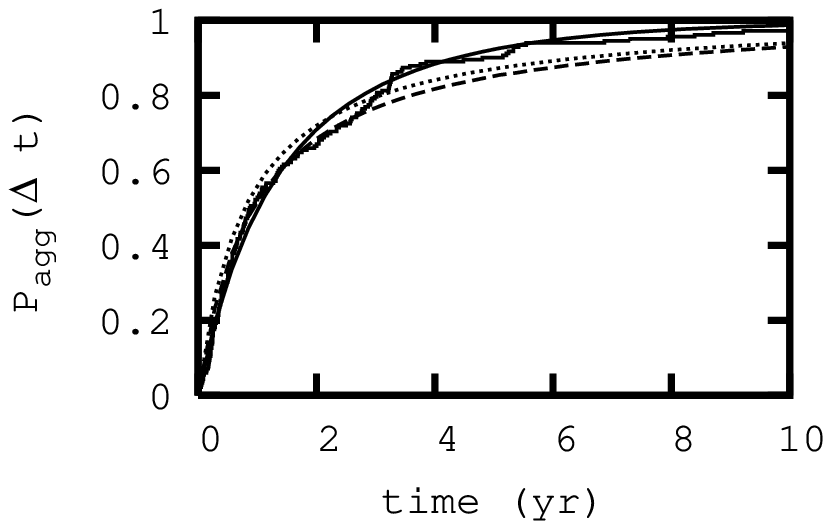}{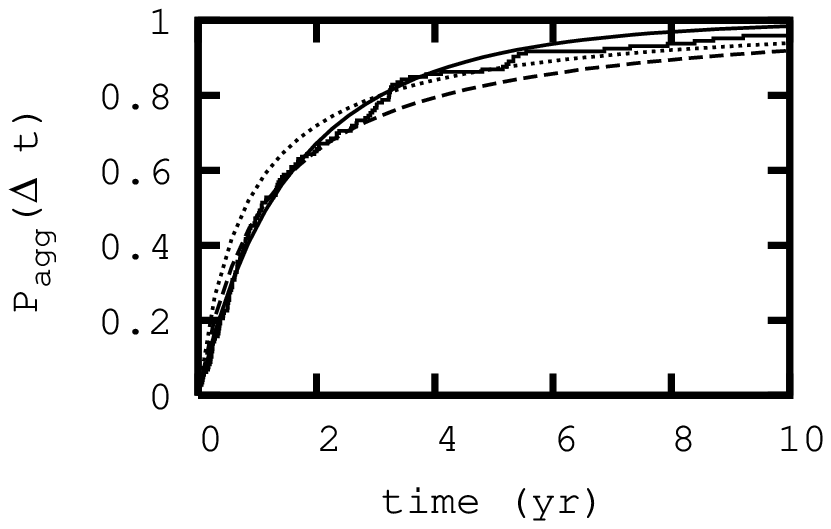}
\caption{Aggregate cumulative waiting-time distribution $P_{\rm agg}(\Delta t)$  
for all pulsars with $N_{\rm g} \geq 2$, 
including ({\em left panel}) and excluding ({\em right panel})
the two quasiperiodic glitchers. The histograms represent
the observational data. The
dotted curves are obtained
by evaluating (\ref{eq:gli11}) with (\ref{eq:gli10}), using
$\langle \lambda \rangle =  1.3_{-0.6}^{+0.7}\,{\rm yr^{-1}}$ ({\em left panel})
and $\langle \lambda \rangle =  1.2_{-0.4}^{+0.5}\,{\rm yr^{-1}}$ ({\em right panel})
extracted from Figure \ref{fig:gli6}. The dashed curves are obtained by
evaluating (\ref{eq:gli11}) with (\ref{eq:gli10}) and adjusting $\langle \lambda \rangle$ to maximize
$P_{\rm KS}$ when fitting $P_{\rm agg}(\Delta t)$.
The solid curves are obtained from (\ref{eq:gli12}),
with $\langle \lambda \rangle = 0.54$ {\rm yr}$^{-1}$,
$\lambda_{\rm min}=0.25$ {\rm yr}$^{-1}$ ({\em left panel})
and $\langle \lambda \rangle = 0.43$ {\rm yr}$^{-1}$,
$\lambda_{\rm min}=0.25$ ({\em right panel}).}
\label{fig:gli7}
\end{figure}

In Figure \ref{fig:gli7}, 
we present the aggregate waiting-time distribution $P_{\rm agg}(\Delta t)$
including ({\em left panel}) and excluding ({\em right panel})
the quasiperiodic glitchers PSR J0537$-$6910 and PSR J0835$-$4510.
The histogram is constructed to include all 182 waiting times in those
objects that have glitched more than once.
The K-S test confirms that the aggregate distribution is fitted poorly
by a single, constant-rate, Poisson distribution of the form
(\ref{eq:gli4}). Furthermore, when we weight (\ref{eq:gli4}) by
the exponential rate distribution (\ref{eq:gli10}),
\footnote{We compute $P_{\rm agg}(\Delta t)$ theoretically as a weighted sum of
independent Poisson processes. In the same way, the waiting-time distribution
for decays observed from a mixture of radioactive isotopes is a sum of
constant-rate Poisson distributions, one per isotope,
weighted by isotopic abundance.} viz.
\begin{eqnarray}
 P_{\rm agg}(\Delta t)
 & = &
 \int_0^{\Delta t} d(\Delta t')
 \int_{0}^{\infty}
 d\lambda' \,
 q(\lambda') p(\lambda',\Delta t'),
 \label{eq:gli11}
\end{eqnarray}
the fit remains poor. For example, the dotted curves in Figure \ref{fig:gli7} are
computed by evaluating (\ref{eq:gli11})
with the mean values
$\langle \lambda \rangle = 1.3_{-0.6}^{+0.7}\,{\rm yr^{-1}}$ ({\em left panel})
and $\langle \lambda \rangle = 1.2_{-0.4}^{+0.5}\,{\rm yr^{-1}}$ ({\em right panel})
extracted from Figure \ref{fig:gli6}. They yield $P_{\rm KS} = 4.3 \times 10^{-2}$
and $2.4 \times 10^{-2}$ respectively. If, instead,
we adjust $\langle \lambda \rangle$ to maximize
$P_{\rm KS}$ while fitting (\ref{eq:gli10}) and (\ref{eq:gli11}) to the observed $P_{\rm agg}(\Delta t)$,
as shown by the dashed curves in Figure \ref{fig:gli7}, we obtain
$\langle \lambda \rangle = 1.1 \,{\rm yr^{-1}}$, $P_{\rm KS} = 0.18$ ({\em left panel})
and $\langle \lambda \rangle = 0.92\,{\rm yr^{-1}}$, $P_{\rm KS} = 0.31$ ({\em right panel}) respectively.

\begin{figure}
\plottwo{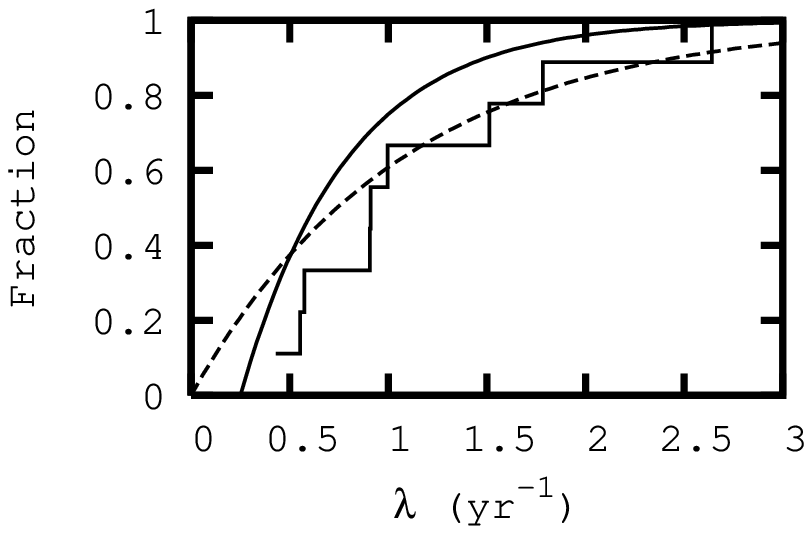}{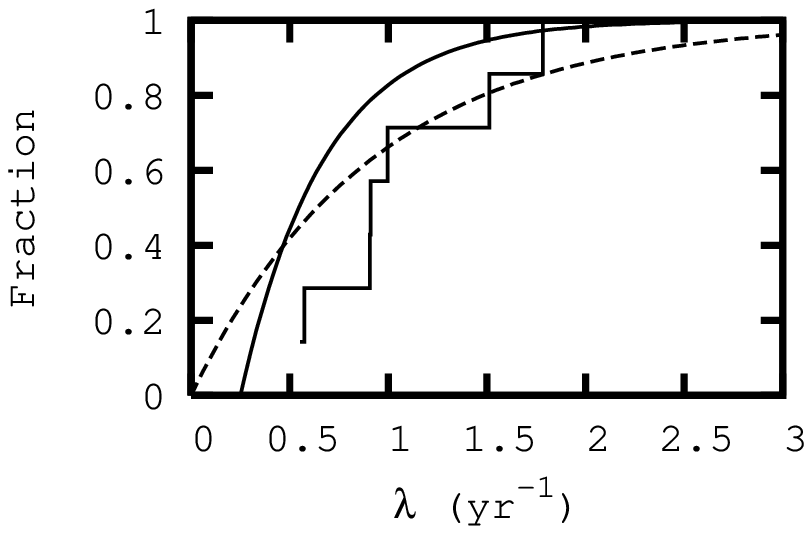}
\caption{Cumulative distribution of mean glitching rate $\lambda$ (measured in {\rm yr}$^{-1}$)
for the nine pulsars that have glitched more than five times
({\em left panel}) and excluding the two quasiperiodic glitchers ({\em right panel}),
showing the observational data ({\em histogram}) and 
the theoretical rate distribution (\ref{eq:14_new})
corresponding to the dashed and solid curves in Figure \ref{fig:gli7}.}
\label{fig:gli6_redo}
\end{figure}

We can exploit the extra information in Figure \ref{fig:gli7}
from objects with $2 \leq N_{\rm g}  \leq 5$ to determine $q(\lambda)$
more accurately.
To do this, we assume a rate
probability density function of the form
\begin{equation}
\label{eq:14_new} q_{\lambda} = \langle \lambda \rangle^{-1}
H(\lambda - \lambda_{\rm min}) \exp[-(\lambda - \lambda_{\rm min})/\langle \lambda \rangle],
\end{equation}
normalize $P(\lambda,\Delta t)$ over the range $[\Delta t_{\rm min}, \Delta t_{\rm max}]$,
and evaluate (\ref{eq:gli11}) to obtain 
\begin{eqnarray}
 P_{\rm agg}(\Delta t)
 & = &
\langle \lambda \rangle^{-1} \int_{\lambda_{\rm min}}^{\infty}
 d\lambda' \, \exp[-(\lambda'-\lambda_{\rm min})/\langle \lambda \rangle] 
\nonumber \\
&  & 
\times \frac{1 - \exp(-\lambda' \Delta t)}{1 - \exp(-\lambda' \Delta t_{\rm max})}
\label{eq:gli12}
\end{eqnarray}
In (\ref{eq:gli12}), we neglect for simplicity the observational bias
introduced by uncertainties in glitch epoch discussed in \S\ref{sec:gli5a};
lacking fuller information, 
we take $\Delta t_{\rm min}=0$ and $\Delta t_{\rm max} = 28.03$ {\rm yr} for all pulsars.
Excellent fits are obtained using (\ref{eq:gli12}).
We find $\langle \lambda \rangle = 0.54$ {\rm yr}$^{-1}$, 
$\lambda_{\rm min}=0.25$ {\rm yr}$^{-1}$, and $P_{\rm KS} = 0.76$ for
all nine pulsars with $N_{\rm g} > 5$,
and $\langle \lambda \rangle = 0.43$ {\rm yr}$^{-1}$,
$\lambda_{\rm min}=0.25$ {\rm yr}$^{-1}$, and $P_{\rm KS} = 0.98$
when the quasiperiodic glitchers are excluded. 
The fits are plotted as solid curves in Figure \ref{fig:gli7}\
({\em left} and {\em right panels} respectively).
We verify the results
by referring back to the measured $q(\lambda)$ distribution.
Substituting the fitted values of $\langle \lambda \rangle$ and
$\lambda_{\rm min}$ into (\ref{eq:14_new}), we obtain
the solid curves in Figure \ref{fig:gli6_redo}, with
$P_{\rm KS} = 0.52$ ({\em left panel})
and $0.25$ ({\em right panel}) respectively. 
Alternatively, the previous fits $\langle \lambda \rangle=1.1$ {\rm yr}$^{-1}$,
$\lambda_{\rm min}=0$ and $\langle \lambda \rangle=0.54$ {\rm yr}$^{-1}$,
$\lambda_{\rm min}=0$ yield
$P_{\rm KS} = 0.72$ ({\em left panel}, dashed curve)
and $0.52$ ({\em right panel}, dashed curve). 

In all cases, $P_{\rm agg} (\Delta t)$
points to the existence of more low-rate objects than 
the $N_{\rm g}>5$ sample in Figure \ref{fig:gli6}
predicts. Specifically, up to $\sim 30$ \% of the population
of glitching pulsars can have $\lambda < 0.25$ {\rm yr}$^{-1}$ while
still reproducing $P_{\rm agg} (\Delta t)$.
We emphasize again that Figure \ref{fig:gli7} constrains $q(\lambda)$ 
more tightly than Figure \ref{fig:gli6}, 
because it contains information about $\Delta t$ from 1.3 times 
as many glitches, including useful information from pulsars which have 
glitched twice.

Undetectable microglitches probably occur between detected glitches
without our knowledge, given that $p(\Delta\nu/\nu)$ is scale invariant.
This effect subtracts from the lower end of the $\Delta t$ distribution 
and adds to the upper end.
We do not correct for it here, because it is hard
to quantify without better statistics.
On two occasions, a pair of glitches occurred on the same date,
once in the same pulsar, and once in different pulsars
\citep{kra05}.
We take $\Delta t = 0$ for these pairs.
Phase connected timing mitigates duty cycle biases,
but it does not eliminate them. 

\subsection{Fluctuation spectrum
 \label{sec:gli5e}}
The temporal fluctuations in a stochastic signal $x(t)$ carry 
independent statistical information about the underlying physical process.
The power spectrum $S(f)$, where $f$ denotes the Fourier frequency,
is related to the temporal autocorrelation function
$G(\tau) = \langle x(t) x(t+\tau) \rangle - \langle x(t) \rangle^2$
(where the average $\langle \dots \rangle$ is performed over $t$ 
for a stationary process)
through the cosine transform
\begin{equation}
 S(f) =
 2\int_0^\infty d\tau \, G(\tau) \cos(2\pi f \tau)~.
 \label{eq:gli13}
\end{equation}

In general, for an avalanche process, the power spectrum depends
jointly on the size, waiting-time, and lifetime distributions
of the avalanches
\citep{jen98}.
For glitches, however, the lifetimes are too short to measure
with current technology (see \S\ref{sec:gli6}).
If, furthermore, we restrict attention to the unit-impulse signal
$x(t) = \sum_i \delta(t-t_i)$,
where $t_i$ denotes the epoch of the $i$-th glitch,
then the sizes drop out of the problem too.
The power spectrum then carries exactly the same information as the
waiting-time distributions 
$P(\lambda,\Delta t)$ and $P_{\rm agg}(\Delta t)$,
with
\begin{equation}
 S(f) \propto 
 \frac{\lambda}{\lambda^2 + (2\pi f)^2}
 \label{eq:gli14}
\end{equation}
for any individual pulsar, and 
\begin{equation}
 S(f) \propto
 \int_{\lambda_{\rm min}}^\infty
 \frac{d\lambda' \, \lambda' q(\lambda')}
  {\lambda'^2 + (2\pi f)^2}
 \label{eq:gli15}
\end{equation}
for the pulsar population in aggregate.

At high frequencies $f\gg\langle\lambda\rangle$, 
equations (\ref{eq:gli14}) and (\ref{eq:gli15}) 
[with $q(\lambda)$ given by (\ref{eq:gli10})]
scale as $f^{-2}$, with $O(f^{-4})$ and $O(f^{-4} \sin 2 f)$ corrections. 
These scalings are modified if the delta function in $x(t)$
is replaced by a nonsingular window function that embodies the
shape of the signal from an individual avalanche.
It will be instructive to revisit this question when it becomes
possible to resolve the lifetimes of individual avalanches,
e.g.\ in single- or giant-pulse timing experiments,

\section{Discussion
 \label{sec:gli6}}
In this paper, we analyze the size and waiting-time distributions
of pulsar glitches, taking advantage of the
enlarged data set produced by the Parkes Multibeam Survey.
We conclude that the data are consistent with the hypothesis
that pulsar glitches arise from an avalanche process.
In each of seven pulsars with $N_{\rm g} > 5$,
the size distribution is consistent with being 
scale invariant across the observed range
of $\Delta\nu$ (up to four decades),
and the waiting-time distribution is consistent
with being Poissonian.
These features are
natural if the system is driven globally at a constant rate
(as the pulsar spins down),
and each glitch corresponds to a locally collective, threshold activated
relaxation of one of the many spatially independent,
metastable stress reservoirs in the system
(e.g. via a vortex unpinning or crust cracking avalanche).
In two pulsars, PSR J0537$-$6910 and PSR J0835$-$4510,
the dynamics may include a second, quasiperiodic component,
comprising $\sim 20\%$ of the events.
The size and waiting-time distributions of the quasiperiodic
component are narrowly peaked,
as expected for rare, system-spanning avalanches, which relax
a large fraction of the total stress accumulated in the system.
This two-component behavior is observed widely in self-organised critical systems,
including experiments on magnetic flux vortices in type II superconductors,
which are closely analogous to neutron star superfluids
\citep{fie95}.

The power-law exponent of the size probability density function
differs from pulsar to pulsar, spanning the range
$-0.13 \leq a \leq 2.4$.
Calculating $a$ theoretically from first principles is a deep problem
which has not yet been solved for other self-organised critical systems,
let alone glitching pulsars,
although some progress has been made on two-dimensional sandpiles using 
renormalization group techniques
\citep{pie94,jen98}.
In the mean-field approximation, which is exact in four or more dimensions, 
theoretical calculations on sandpiles (and other systems in their
universality class) yield $a = 1.5$,
whereas three-dimensional cellular automata output $a=1.3$
\citep{jen98}.

\begin{figure}
\includegraphics[angle=270,scale=0.9]{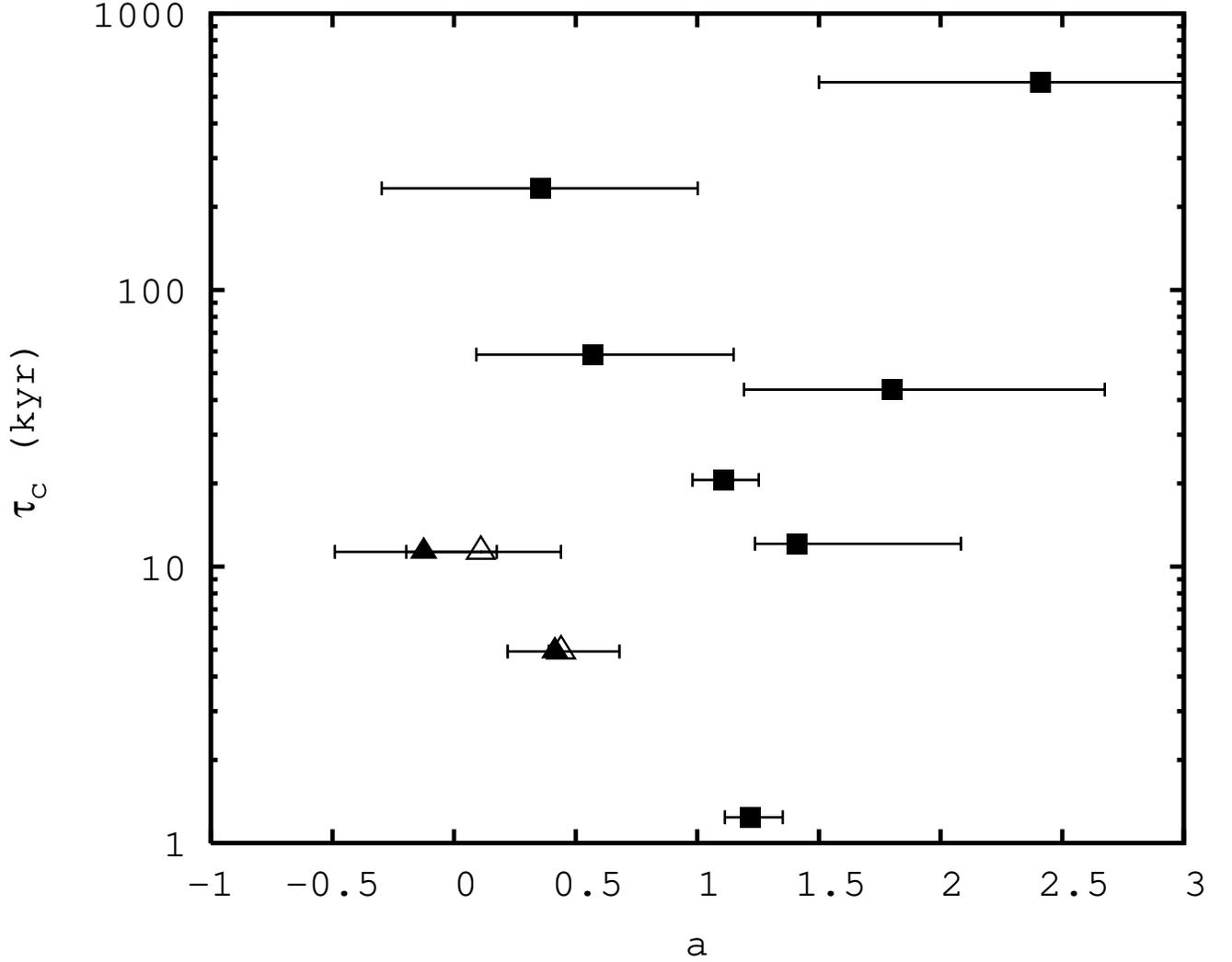}
\caption{Spin-down age $\tau_c$ (in {\rm kyr}) versus
power-law exponent $a$ for the glitch size distribution.
The error bars indicate the 1-$\sigma$ range of allowable
fits according to the K-S test. Systematic differences between
$\tau_c$ and true age are not quantified here. Solid (open)
triangles symbolize quasiperiodic glitchers with $N_{\rm g} > 5$,
to which we fit a two-component (one-component) $P(\Delta \nu/\nu)$, as in
Table \ref{tab:gli3} (\ref{tab:gli2}). Squares symbolize aperiodic glitchers with $N_{\rm g} > 5$,
to which we fit a power-law $P(\Delta \nu/\nu)$, as in 
Table \ref{tab:gli2}.}
\label{fig:gli9}
\end{figure}

The size distribution transmits two important lessons
concerning the microphysics of glitches.
First, the fact that $a$ differs from pulsar to pulsar implies
that the strength and level of conservation of the local 
(e.g.\ pinning and intervortex) forces also differs
\citep{ola92,fie95}.
By contrast, in equilibrium critical systems like ferromagnets,
$a$ depends only on the dimensionality of the system and its
order parameter and is therefore universal.
Second, except for the two pulsars which show quasiperiodicity,
$a$ appears to vary smoothly with spin-down age,
with $a\approx 1.2$ for the youngest pulsars (e.g.\ the Crab).
Figure \ref{fig:gli9} depicts the trend between $a$ and $\tau_{\rm c}$.
It is suggestive; after all, local pinning forces do depend on
temperature and hence $\tau_{\rm c}$.
Interestingly, however, there is no clear trend between $a$ and $\nu$,
even though the mean vortex spacing (and hence intervortex force)
is proportional to $\nu^{1/2}$.
It will pay to study these trends more thoroughly
as more glitch data is collected.

An avalanche process predicts a specific relation between
the distributions of glitch sizes $\Delta\nu$ and lifetimes $T$
(as opposed to waiting times $\Delta t$).
Specifically, in a self-organized critical state,
the lifetime probability density function is also a power law,
$p(T) \propto T^{-b}$,
with
\begin{equation}
 b = 1 + (a-1)\gamma_2 / \gamma_3~.
 \label{eq:gli16}
\end{equation}
The constants $\gamma_2$ and $\gamma_3$ are defined such that the cardinality 
of an avalanche scales with its linear extent ($L$) as $L^{\gamma_2}$
and its lifetime (i.e.\ duration) scales as $L^{\gamma_3}$
\citep{jen98}.
Both $\gamma_2$ and $\gamma_3$ depend on the effective dimensionality of the 
local forces
and can be calculated numerically using a cellular automaton.
In two dimensions, avalanches are compact, not fractal,
and one has $\gamma_2=2$;
in three dimensions, one has $2 < \gamma_2 < 3$.
At present, radio timing experiments cannot measure $T$;
most glitches are detected as unresolved, discontinuous, spin-up events 
with $T < 120\,{\rm s}$ \citep{mcc90}.
\footnote{In the Crab, some spin-up events seem to be resolved,
e.g.\ at epochs MJD 50260 ($T \approx 0.5\,{\rm d}$)
and MJD 50489 (secondary spin up, $T \approx 2\,{\rm d}$)
\citep{won01}.
If these are rare but otherwise standard glitches originating from the long-$T$
tail of the lifetime distribution, it is puzzling that other, shorter,
but still resolved (and presumably more common) spin-up events 
are not observed, e.g.\ with $T\sim 0.1\,{\rm d}$ or $0.01\,{\rm d}$.
Alternatively, the events at MJD 50620 and MJD 50489
may have been triggered by a different physical mechanism.}
In the future, however, single- and/or giant-pulse timing experiments
with more sensitive instruments (e.g.\ the Square Kilometer Array) 
will test this prediction.
If confirmed, it will independently corroborate the avalanche hypothesis.

The mean glitching rates of the nine pulsars studied here
are fairly narrowly distributed,
spanning the range
$0.35\,{\rm yr^{-1}} \leq \lambda \leq 2.6\,{\rm yr^{-1}}$.
The probability density function for $\lambda$ is adequately
fitted by an exponential, as for solar flare avalanches \citep{whe00},
with $\langle\lambda\rangle = 1.3_{-0.6}^{+0.7}$ {\rm yr}$^{-1}$,
or by an exponential with a lower cutoff, at $\lambda_{\rm min} \approx 0.25$ {\rm yr}$^{-1}$.
A theoretical derivation of $\langle\lambda\rangle$ from first principles
is currently lacking,
although estimates of how long it takes to crack the crust locally
predict reasonable rates,
if the critical strain angle approaches that of imperfect terrestrial metals
\citep{alp96,mid06}.

\begin{figure}
\begin{center}
\includegraphics[angle=270,scale=0.9]{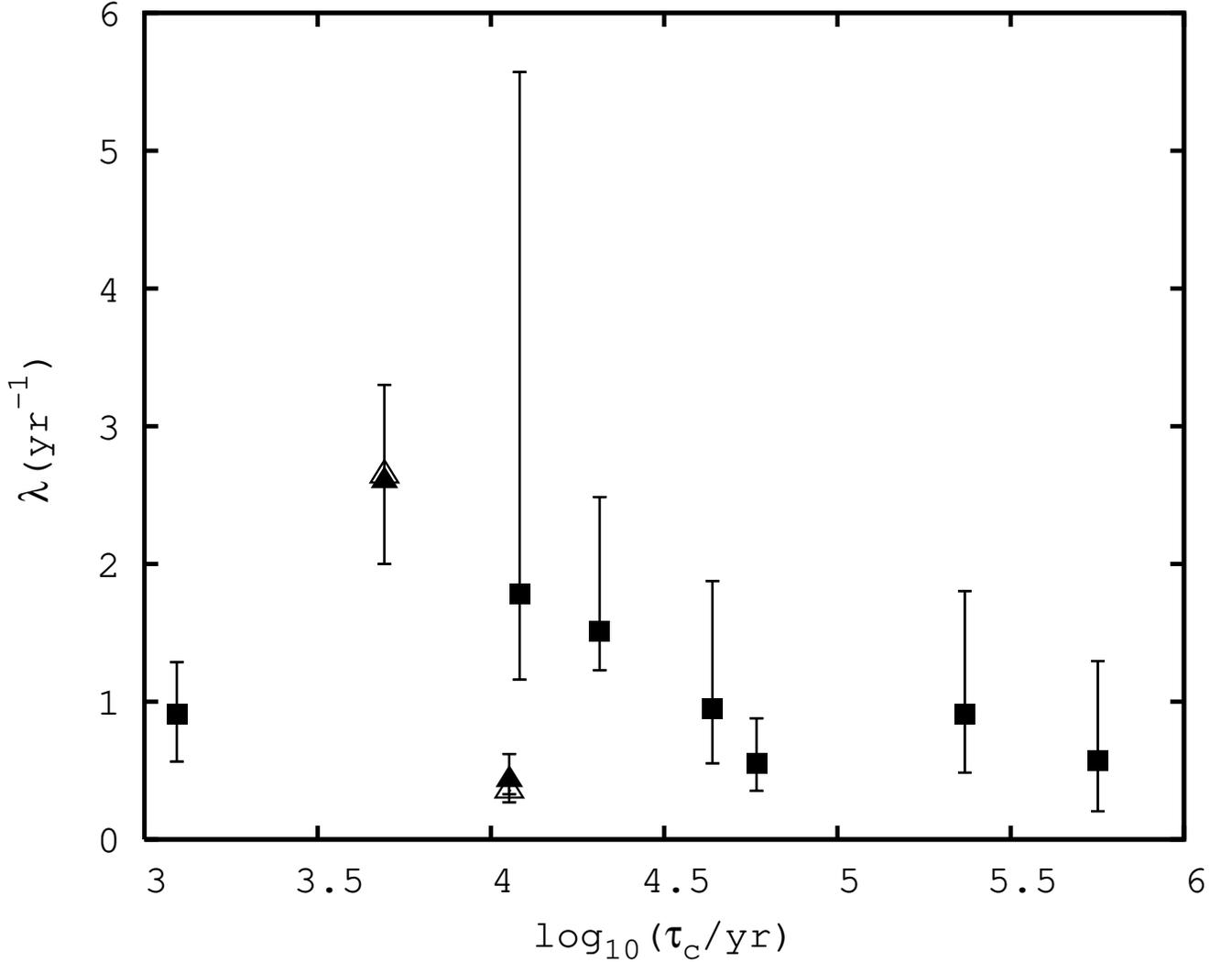}
\caption{Mean glitching rate $\lambda$ (in {\rm yr}$^{-1}$)
versus spin-down age $\tau_c$ (in {\rm kyr}).
The error bars indicate the 1-$\sigma$ range of allowable
fits according to the K-S test. Systematic differences between
$\tau_c$ and true age are not quantified
here. Solid (open) triangles symbolize quasiperiodic 
glitchers with $N_{\rm g} > 5$,
to which we fit a two-component (one-component) $P(\lambda,\Delta t)$, as in
Table \ref{tab:gli5} (\ref{tab:gli4}). Squares symbolize aperiodic glitchers with $N_{\rm g} > 5$,
to which we fit a Poissonian $P(\lambda,\Delta t)$ as in Table \ref{tab:gli4}.}
\label{fig:gli10}
\end{center}
\end{figure}

Figure \ref{fig:gli10} plots $\lambda$ versus $\tau_c$ for the nine
pulsars examined individually in this paper. There is no
significant trend. The data are consistent with the notion that
old pulsars glitch less frequently than young pulsars
\citep{she96}, but they are equally consistent with the notion
that the glitching rate is independent of age.

Many authors have searched for a correlation between waiting time
and the size of the next glitch.
Such a correlation appears to be absent from the data,
e.g.\ Figure 17 in \citet{wan00}
and Figure 10 in \citet{won01}.
At first blush, this is surprising: the vortex unpinning and
crust fracture paradigms, which are driven by the accumulation
of differential rotation and mechanical stress respectively,
seem to be natural candidates for a `reservoir effect'.
Avalanche dynamics resolves this apparent paradox.
The reservoir effect does operate locally, 
but the star contains many reservoirs,
insulated from each other by relaxed zones,
whose storage capacities evolve stochastically in response to
the slow driver and avalanche history.
During a glitch, a single reservoir
(often small but sometimes large) relaxes at random via an avalanche,
releasing its stored $\Delta\nu$
(and destabilizing neighboring reservoirs in preparation for
the next glitch).
Some of the $\Delta\nu$ is accumulated since the previous glitch, 
but the remainder is `borrowed' from earlier epochs,
when some other reservoir relaxed instead.
All self-organized critical systems share these dynamics;
the waiting time is uncorrelated with the size of the next avalanche
\citep{jen98}.
The only exceptions are large, system-spanning avalanches,
which always have roughly the same sizes and waiting times,
and which account for $\sim 20\%$ of the 
glitches in PSR J0537$-$6910 and PSR J0835$-$4510.

A corollary of the previous paragraph is that the total $\Delta\nu$
released in glitches up to some epoch is less than the total crust-superfluid
differential rotation accumulated since that epoch, viz.
\begin{equation}
 \sum_{i=1}^{N_{\rm g}} \Delta\nu_i
 \leq 
 \epsilon |\dot{\nu}|  \sum_{i=1}^{N_{\rm g}} \Delta t_{i},
 \label{eq:gli17}
\end{equation}
where $\epsilon\dot{\nu}$ is the relative angular acceleration of the
crust and superfluid due to electromagnetic spin down.
The `staircase' described by (\ref{eq:gli17}) has been noted previously 
\citep{she96,lyn00}, 
both in quasiperiodic glitchers like PSR J0537$-$6910
[e.g.\ Figure 8 in \citet{mid06}],
where the reservoir effect is obvious,
and in Poisson glitchers like PSR J0534$+$2200,
[e.g.\ Figure 12 in \citet{won01}],
where the trend is more subtle because it reverts to the mean
over long times, not after every glitch.
Upon dividing (\ref{eq:gli17}) by $N_{\rm g}$,
and averaging over long times,
the inequality becomes an equality (provided there is no secular
accumulation of differential rotation in the system)
and we recover (\ref{eq:gli5}). 

It is fundamentally impossible to measure $\epsilon$
for individual pulsars with current data, because
$\langle \Delta \nu \rangle$ is dominated by large
(and therefore rare) glitches for $a < 2$. It is therefore wrong
to assume stationarity over a typical, $40$-yr observation
interval. Consequently, we are prompted to reassess
the familiar correlation between activity and spin-down age
\citep{she96}.
Our definition of $\epsilon\dot{\nu}$ is identical to 
$\dot{\nu}_{\rm glitch}$ in \citet{lyn00}
(but for individual objects, not in aggregate)
and $A_{\rm g}$ in \citet{won01}.
It is closely related to the original activity parameter
defined by \citet{mck90},
which equals $N_{\rm g} \nu^{-1} \epsilon |\dot{\nu}|$.
For PSR J0358$+$5413, we measure
$\epsilon\leq 7 \times 10^{-5}$, lower than
the {\it aggregate} value $0.017\pm0.002$
measured by \citet{lyn00} for objects with 
$\tau_{\rm c} > 10\,{\rm kyr}$ (binned by semi-decades in $\dot{\nu}$).
\footnote{The aggregate value $\dot{\nu}_{\rm glitch}$
\citep{lyn00}, binned over semi-decades in $\dot{\nu}$, effectively averages
together different pulsars. While this approach reduces the formal
error bar on $\dot{\nu}_{\rm glitch}$, its physical interpretation
is less straightforward, given the likelihood that $\epsilon$
is different in different pulsars.} 
Interpreted in terms of the vortex unpinning model,
this result suggests that $0.007$--$2$ \% of the angular momentum outflow
during spin down may be stored in metastable reservoirs on average
over time. On the other hand, five other objects have
$0.04 \leq \epsilon_{\rm max} \leq 0.8$, under the questionable
assumption that the maximum physical size is 
$\Delta \nu_{\rm upper} = 2 \times 10^{-4}$
in all pulsars. Our data are therefore inadequate to update usefully
the value $A_{\rm g}/|\dot{\nu}| = 1\times 10^{-5}$ measured by \citet{won01}
for PSR J0534$+$2200.

In the context of vortex unpinning, it has been argued that
the aggregated $\epsilon$ measured by \citet{lyn00}
partly corroborates 
the hypothesis that younger pulsars are still in the process of
forming their capacitive elements,
e.g.\ by creating pinning centers through crust fracture,
while older pulsars have mostly completed the task
\citep{alp96,won01}.
However, the full picture is more complicated.
Vela's quasiperiodic avalanches point to a richly
connected network of reservoirs
\citep{alp96},
yet its aggregated value $\dot{\nu}_{\rm glitch}$
is relatively low.
On the other hand, the other quasiperiodic glitcher,
PSR J0537$-$6910, is relatively young 
($\tau_{\rm c} = 4.9\,{\rm kyr}$);
how did it form a richly connected reservoir network so quickly?
And, if its network is so richly connected, why is its 
aggregated $\dot{\nu}_{\rm glitch}$ value so low? Likewise,
PSR J0358$+$5413 is the oldest object in
the sample ($\tau_{\rm c} = 560 \,{\rm kyr}$),
yet its $\epsilon$ value arguably points to a dearth
of capacitive elements, characteristic of a young object.
There are no obvious grounds (e.g.\ quasiperiodicity) 
on which to treat PSR J0358$+$5413 as exceptional.

Do all pulsars glitch eventually?
It has been speculated in the past that there is something special
physically about the minority of pulsars that do glitch.
While it is impossible to reject this hypothesis unequivocally
with the data at hand,
the results presented here suggest
that all pulsars are capable of glitching.
However, most do so infrequently (low $\lambda$)
and hence have not been detected
during the last four decades of timing experiments.
We find that up to $\sim 30 \%$ of the pulsar population
can glitch at rates lower than $\lambda_{\rm min}= 0.25$ {\rm yr}$^{-1}$
and still conform with the measured aggregate waiting-time distribution.

Once verified, the claimed Poissonian nature of the glitch mechanism can be
invoked to exclude broad classes of glitch theories,
e.g.\ those that rely on `defects' or `turbulence' at special locations
(like the pole), 
or that involve a pair of dependent events
(A.\ Martin, private communication).
It is important to interpret aftershocks carefully in this light
\citep{won01}.
In self-organized critical systems,
the excess number of avalanches following a large avalanche
(over and above the Poissonian baseline following a small avalanche)
scales inversely with the time elapsed, a property known as 
Omori's law for earthquakes
\citep{jen98}.

In this paper, we do not analyze post-glitch relaxation times 
and glitch-activated changes in $\dot{\nu}$ in the context of
avalanche processes,
e.g.\ the correlation between $\Delta\dot{\nu}$ and the transient
component of $\Delta\nu$
\citep{won01}.
We also assume implicitly that the quantized superfluid vortices 
in the vortex unpinning model are organized in a rectilinear array,
even though recent work suggests that meridional circulation
destabilizes the array and converts it into a turbulent tangle
\citep{per05,per06}. Further study of
these matters is deferred to future work.

We acknowledge the computer time and system support
supplied by the Australian Partnership for
Advanced Computation (APAC) and the
Victorian Partnership for
Advanced Computation (VPAC).
We thank Andre Trosky for illuminating conversations on
self-organized criticality and cellular automata.
This research was supported by
a postgraduate scholarship from the University of Melbourne.
It makes use of the Australia Telescope National Facility 
Pulsar Catalogue
\citep{man05}, 
which can be accessed on-line at
{\tt http://www.atnf.csiro.au/research/pulsar/psrcat}.


\newpage

\renewcommand{\thefootnote}{\alph{footnote}}

\begin{deluxetable}{cccclcc}
\tablewidth{0pt}
\tablecaption{Parameters of pulsar glitches}
\tablehead{PSR J & $ N_{\rm g}$ &	Epoch	& $\Delta \nu/\nu$&  Ref.\\
&  & (MJD) & $(10^{-9})$ & }

\startdata
0142+61 \phantom{12}&		1	&	51141	&	650	&	6 \\ \tableline
0157+6212	&	1	&	48504	&	2.46	&	6 \\ \tableline
0358+5413	&   	6	&	46077(2)	&	5.5(1)	&	1 \\ 
		&   	&	46497(4)	&	4368(1)	&	1 \\
	&   	&	51673(15)	&	0.04(2)	&	31 \\
	&   		&	51965(14)	&	0.030(2)	&	31 \\
	&   	&	52941(9)	&	0.04(1)	&	31 \\
	&   	&	53216(11)	&	0.10(2)	&	31 \\ \tableline
0528+2200	&		3	&	42057	&	1.3	&	1 \\
	&			&	52289	&	1.46	&	31 \\
	&			&	53379	&	0.17	&	31 \\ \tableline
0534+2200	&    	26	&	40493.4(1)	&	4(2)	&	1 \\
	&			&	41163(1)	&	2.2(1)	&	24 \\
	&			&	41250(1)	&	2(1)	&	23 \\
	&			&	42448(1)	&	44.0(6)	&	1 \\
	&			&	43023(1)	&	1.1(1)	&	24 \\
	&			&	43768(1)	&	2.8(1)	&	24 \\
	&			&	46664.42(5)	&	4.1(1)	&	1 \\
	&			&	47768.4(2)	&	85.0(4)	&	1 \\
	&			&	48945.5(2)	&	4.5(7)	&	1 \\
	&			&	50020.6(3)	&	2.7(7)	&	4 \\
	&			&	50259.93(2)	&	22(1)	&	4 \\
	&			&	50459.1(5)	&	7.67(3)	&	4 \\
	&			&	50489(2)	&	6.67	&	4 \\
	&			&	50812.9(1)	&	8.67(2)	&	4 \\
	&			&	51452.3(1)	&	9.67(2)	&	4 \\
	&			&	51741(5)	&	24(5)	&	25 \\
	&			&	51805.03(3)	&	3.3(2)	&	29 \\
	&			&	52083.969(2)	&	23.6(6)	&	29 \\
	&			&	52146.757(9)	&	8(1)	&	29 \\
	&			&	52498.22(6)	&	2.6(2)	&	29 \\
	&			&	52587.84(3)	&	1.1(2)	&	29 \\
	&			&	53067.059(1)	&	210(1)	&	29 \\
	&			&	53254.039(1)	&	4.84(8)	&	29 \\
	&			&	53331(1)	&	n/a	&	29 \\
	&			&	53463.72(3)	&	n/a	&	29 \\
	&			&	53476.7(8)	&	n/a	&	29 \\ \tableline
0537$-$6910	&   	23	&	51285(8.6)	&	681(65)	&	33 \\
	&			&	51568(6.8)	&	449(8)	&	33 \\
	&			&	51711(6.7)	&	315(9)	&	33 \\
	&			&	51826(7.1)	&	140(7)	&	33 \\
	&			&	51880(5.5)	&	141(20)	&	33 \\
	&			&	51959(4.9)	&	456(46)	&	33 \\
	&			&	52171(8.3)	&	185(6)	&	33 \\
	&			&	52242(7.8)	&	427(6)	&	33 \\
	&			&	52386(5.7)	&	168(20)	&	33 \\
	&			&	52453(6.9)	&	217(30)	&	33 \\
	&			&	52546(6.2)	&	421(18)	&	33 \\
	&			&	52740(5.3)	&	144(6)	&	33\\
	&			&	52819(3.6)	&	256(16)	&	33 \\
	&			&	52887(4.5)	&	234(23)	&	33 \\
	&			&	53014(9.5)	&	338(10)	&	33 \\
	&			&	53125(2.8)	&	18(14)	&	33 \\
	&			&	53145(2.7)	&	392(8)	&	33 \\
	&			&	53288(2.4)	&	395(10	&	33 \\
	&			&	53446(1.7)	&	259(16)	&	33 \\
	&			&	53551(4.4)	&	322(26)	&	33 \\
	&			&	53699(3.9)	&	402(8)	&	33 \\
	&			&	53860(1.5)	&	236(20)	&	33 \\
	&			&	53951(1.5)	&	18(20)	&	33 \\ \tableline
0540$-$6919			&	1	&	51325	&	1.9	&	8 \\ \tableline
0601$-$0527	&	1	&	51662	&	0.19	&	28,29 \\ \tableline
0614+2229	&			1	&	51339	&	n/a	&	28,29 \\ \tableline
0631+1036	&     	9	&	50185.711(6)	&	5.1(1)	&	28,29 \\
	&			&	50479.74(7)	&	3.7(2)	&	29 \\
	&			&	50608.246(2)	&	57.9(3)	&	29 \\
	&			&	50730(2)	&	1662.8(4)	&	29 \\
	&			&	51911.133(8)	&	1.33(8)	&	29 \\
	&			&	52852.586(2)	&	17.4(2)	&	29 \\
	&			&	53228.387(2)	&	1.9(1)	&	29 \\
	&			&	53359.27(1)	&	1.9(3)	&	29 \\ 
	&			&	53621(2)	&	n/a	&	29 \\ \tableline
0659+1414			&	2	&	50185	&	0.39	&	28,29 \\ 
	&			&	51039	&	1.4	&	29 \\ \tableline
0729$-$1448	&		1	&	52149.6	&	31	&	28,29 \\ \tableline
0742$-$2822	&		5	&	n/a	&	n/a	&	28 \\
	&				&	51770	&	1	&	31 \\
	&				&	52027	&	2.1	&	31 \\
	&				&	53090.2	&	2.9	&	31 \\
	&				&	53469.7	&	1.1	&	31 \\ \tableline
0745$-$5353	&		1	&	n/a	&	n/a	&	 \\ \tableline
0758$-$1528	&		1	&	49948	&	0.13	&	28,29 \\ \tableline
0826+2637	&		1	&	n/a	&	n/a	&	28 \\ \tableline
0835$-$4510	&     	17	&	40280(4)	&	2340(10)	&	1 \\
	&			&	41192(8)	&	2050(30)	&	1 \\
	&			&	41312(4)	&	12	&	26 \\
	&			&	42683(3)	&	1990(10)	&	1 \\
	&			&	43693(12)	&	3060(60)	&	1 \\
	&			&	44888.0707(4)	&	1145(3)	&	1 \\
	&			&	45192(5)	&	2050(10)	&	1 \\
	&			&	46257.2284(2)	&	1601(1)	&	1 \\
	&			&	47519.803(8)	&	1807.1(8)	&	1 \\
	&			&	48457.382(10)	&	2715(2)	&	1 \\
	&			&	48550(1)	&	5.6	&	3 \\
	&			&	49559.057(2)	&	835(2)	&	1 \\
	&			&	49591.158(2)	&	199(2)	&	1 \\
	&			&	50369.345(2)	&	2110(17)	&	3 \\
	&			&	51559.345(5)	&	3120	&	27 \\
	&			&	53195.09(5)	&	2100	&	30 \\
	&			&	53959.93(2)	&	2620	&	34 \\ \tableline
1016$-$5857	&	1	&	52550	&	n/a	&	29,32 \\ \tableline
1048$-$5832		&	4	&	48944	&	19	&	3 \\
	&			&	49034	&	3000	&	3 \\
	&			&	50788	&	769	&	3 \\
	&			&	52733	&	n/a	&	29 \\ \tableline
1105$-$6107	&		2	&	50417	&	279.7	&	3 \\
	&				&	50610	&	2.1	&	3 \\ \tableline
1112$-$6103	&		1	&	51513	&	n/a	&	29 \\ \tableline
1119$-$6127	&		2	&	51398	&	4.4	&	9 \\
	&			&	53300	&	100	&	29 \\ \tableline
1123$-$6259		&	1	&	49705.87	&	749.31	&	3 \\ \tableline
1141$-$3322		&	1	&	50551	&	0.7	&	28,29 \\ \tableline
1302$-$6350		&	1	&	50690.7	&	3.2	&	10 \\ \tableline
1328$-$4357		&	1	&	43590	&	1.16	&	1 \\ \tableline
1341$-$6220		&	12	&	47989(24)	&	1507(1)	&	3 \\
	&		&	48453(12)	&	24.2(9)	&	3 \\
	&		&	48645(10)	&	990(3)	&	3 \\
	&		&	49134(22)	&	10(2)	&	3 \\
	&		&	49363(130)	&	142(21)	&	3 \\
	&		&	49523(17)	&	33(3)	&	3 \\
	&		&	49766(2)	&	11(1)	&	3 \\
	&		&	49904(16)	&	16(7)	&	3 \\
	&		&	50008(16)	&	1636(13)	&	3 \\
	&		&	50321.7(6)	&	27(1)	&	3 \\
	&		&	50528.9(8)	&	20(4)	&	3 \\
	&		&	50683(13)	&	703(4)	&	3 \\ \tableline
1357$-$6429		&	1	&	52021	&	2425	&	11 \\ \tableline
1401$-$6357		&	1	&	48305	&	2.49	&	28,29 \\ \tableline
1413$-$6141		&	1	&	n/a	&	n/a	&	29 \\ \tableline
1437$-$6146		&	1	&	51614	&	n/a	&	29,32 \\ \tableline
1509+5531		&	1	&	41732	&	0.22	&	11 \\ \tableline
1532+2745		&	1	&	n/a	&	n/a	&	28 \\ \tableline
1539$-$5626		&	1	&	48165	&	2790.8	&	1 \\ \tableline
1603$-$2531		&	1	&	n/a	&	n/a	&	 \\ \tableline
1614$-$5048		&	1	&	49803	&	6460	&	3 \\ \tableline
1617$-$5055		&	1	&	49960	&	600	&	13 \\ \tableline
1644$-$4559		&	3	&	43390	&	191	&	1 \\
	&		&	46453	&	803.6	&	1 \\
	&		&	47589	&	1.61	&	1 \\ \tableline
1705$-$1906		&	1	&	48888	&	0.44	&	28,29 \\ \tableline
1705$-$3423		&	2	&	50060	&	n/a	&	28,29 \\
	&		&	51940	&	0.6	&	28,29 \\ \tableline
1708$-$4008		&	2	&	51459	&	620	&	14 \\
	&		&	52014	&	140	&	15 \\ \tableline
1709$-$4429	&	1	&	48780	&	2050	&	1 \\ \tableline
1717$-$3425	&	1	&	49868	&	n/a	&	29 \\ \tableline
1720$-$1633	&	1	&	n/a	&	n/a	&	 \\ \tableline
1721$-$3532	&	1	&	49969.7	&	8	&	29 \\ \tableline
1726$-$3530	&	1	&	n/a	&	n/a	&	29 \\ \tableline
1730$-$3350	&		2	&	47990	&	3080	&	1 \\
	&		&	52139	&	3190	&	29 \\ \tableline
1731$-$4744	&	2	&	49397.3	&	139.2	&	3 \\
	&			&	50703	&	3.1	&	3 \\ \tableline
1737$-$3137	&	2	&	51553	&	4	&	29 \\
	&			&	53052.8	&	236	&	29 \\ \tableline
1739$-$2903	&	1	&	46956	&	3.09	&	1 \\
1739$-$3131	&	1	&	n/a	&	n/a	&	28 \\ \tableline
1740+1311	&	1	&	n/a	&	n/a	&	28 \\ \tableline
1740$-$3015	&	30	&	47003(50)	&	420(20)	&	1 \\
	&	&	47281(2)	&	33(5)	&	1 \\
	&	&	47332(16)	&	7(5)	&	1 \\
	&	&	47458(2)	&	30(8)	&	1 \\
	&	&	47670.2(2)	&	600.9(6)	&	1 \\
	&	&	48149(2)	&	4(2)	&	29 \\
	&	&	48186(6)	&	642(16)	&	5 \\
	&	&	48218(2)	&	48(10)	&	5 \\
	&	&	48431(2)	&	15.7(5)	&	5 \\
	&	&	49046(4)	&	9.1(2)	&	1 \\
	&	&	49239(2)	&	169.7(2)	&	1 \\
	&	&	49451.7(4)	&	9.5(5)	&	2 \\
	&	&	49543.9(8)	&	3(6)	&	2 \\
	&	&	50574.5497(4)	&	439.3(2)	&	2 \\
	&	&	50941.6182(2)	&	1443(3)	&	2 \\
	&	&	51334(2)	&	1.1(6)	&	29 \\
	&	&	51685(24)	&	0.7(4)	&	31 \\
	&	&	51822(7)	&	0.8(3)	&	31 \\
	&	&	52007(6)	&	0.7(1)	&	31 \\
	&	&	52235(2)	&	42.1(9)	&	31 \\
	&	&	52240.2(2)	&	5(1)	&	29 \\
	&	&	52266.8(2)	&	14.3(7)	&	29 \\
	&	&	52271(2)	&	444(5)	&	31 \\
	&	&	52344(2)	&	220.6(9)	&	31 \\
	&	&	52603(5)	&	1.5(1)	&	31 \\
	&	&	52759(5)	&	1.6(3)	&	31 \\
	&	&	52859(2)	&	17.6(3)	&	31 \\
	&	&	52943.5(2)	&	22.1(4)	&	31 \\
	&	&	53023.512(2)	&	1850.0(8)	&	29 \\
	&	&	53741(2)	&	n/a	&	29 \\ \tableline
1743$-$3150	&	1	&	49553	&	1.6	&	28,29 \\ \tableline
1755$-$2534	&	1	&	52170	&	n/a	&	29 \\ \tableline
1759$-$2205	&	1	&	51800	&	28	&	29 \\ \tableline
1801$-$0357	&	1	&	48016	&	2.9	&	2 \\ \tableline
1801$-$2304	&	9	&	46907(40)	&	200(30)	&	1 \\
	&	&	47855(50)	&	231.2(9)	&	1 \\
	&	&	48454(3)	&	347.68(8)	&	1 \\
	&	&	49709(32)	&	64(2)	&	5 \\
	&	&	50055(4)	&	22.6(9)	&	2 \\
	&	&	50363.414(4)	&	80.6(6)	&	2 \\
	&	&	50938(2)	&	4(1)	&	29 \\
	&	&	52126(100)	&	651(3)	&	29 \\
	&	&	53356(100)	&	499(4)	&	29 \\ \tableline
1801$-$2451	& 	5	&	49476	&	1988	&	3 \\
	&	&	50651	&	1247	&	3 \\
	&	&	52567	&	n/a	&	29 \\
	&	&	52791	&	n/a	&	29 \\
	&	&	53030.51	&	16.1	&	29 \\ \tableline
1803$-$2137	&	4	&	48245	&	4075	&	1 \\
	&	&	50269.4	&	5.3	&	2 \\
	&	&	50765	&	3185	&	3 \\
	&	&	50765	&	27	&	3 \\
	&	&	53429	&	3943	&	29 \\ \tableline
1806$-$2125	&	1	&	51063	&	15615	&	16 \\ \tableline
1809$-$1917	&	1	&	53250	&	1629.1	&	29 \\ \tableline
1812$-$1718	&	2	&	49926	&	1.6	&	28,29 \\
	&	&	53105.68	&	14.7	&	28,29 \\ \tableline
1814$-$1744	&	5	&	51384	&	9	&	29 \\
	&			&	51700	&	5	&	31 \\
	&			&	52094.96	&	27	&	29 \\
	&			&	52117	&	33	&	31 \\
	&			&	53302	&	7	&	31 \\ \tableline
1824$-$1118	&	1	&	52402	&	1.3	&	28,29 \\ \tableline
1824$-$2452	&	1	&	51980	&	0.0095	&	17 \\ \tableline
1825$-$0935	&	8	&	49615(8)	&	0.2(1)	&	21 \\
	&			&	49857(8)	&	12.6	&	22 \\
	&			&	49940(2)	&	5.21(7)	&	21 \\
	&			&	50557(6)	&	12.6(2)	&	21 \\
	&			&	51060(8)	&	20	&	22 \\
	&			&	51879(8)	&	31.4(2)	&	22 \\
	&			&	52058(2)	&	29(1)	&	29 \\
	&			&	52802.6(2)	&	1.8(7)	&	29 \\ \tableline
1826$-$1334	&	3	&	46507	&	2700	&	1 \\
	&			&	49014	&	3060	&	1 \\
	&			&	53738	&	n/a	&	29 \\ \tableline
1827$-$0958	&	1	&	n/a	&	n/a	&	28 \\ \tableline
1833$-$0559	&	1	&	52200	&	n/a	&	29 \\ \tableline
1833$-$0827	&	1	&	48041	&	1864.8	&	1 \\ \tableline
1835$-$1106	&	1	&	52265	&	27	&	28,29 \\ \tableline
1838$-$0453	&	1	&	52000	&	26	&	29 \\ \tableline
1841$-$0425	&	1	&	53356	&	578.5	&	29 \\ \tableline
1844$-$0538	&	2	&	47438	&	0.8	&	29 \\
	&			&	47955	&	0.5	&	29 \\ \tableline
1845$-$0316	&	1	&	52212.9	&	30	&	29 \\ \tableline
1856+0113			&	1	&	n/a	&	n/a	&	28 \\ \tableline
1901+0716	&			1	&	46859	&	30	&	1 \\ \tableline
1902+0615	&			4	&	48645.11	&	0.45	&	29 \\
	&			&	49441	&	0.23	&	29 \\
	&			&	50311	&	0.31	&	29 \\
	&			&	51165.9	&	0.47	&	29 \\ \tableline
1903+0135	&	1	&	48634	&	n/a	&	28,29 \\ \tableline
1905$-$0056	&	2	&	49385	&	n/a	&	28,29 \\
	&			&	49695.3	&	0.21	&	28,29 \\ \tableline
1908+0909	&	2	&	52240	&	11.8	&	29 \\
	&			&	53340	&	1.7	&	29 \\ \tableline
1909+0007	&	1	&	49491.9	&	0.72	&	29 \\ \tableline
1910+0358	&	1	&	52331	&	1.4	&	29 \\ \tableline
1910$-$0309	&	3	&	48241	&	0.6	&	2 \\
	&			&	49219.85	&	1.84	&	2 \\
	&			&	53232.75	&	2.66	&	29 \\ \tableline
1913+0446	&	1	&	53500	&	n/a	&	29 \\ \tableline
1918+1444	&	1	&	52285	&	2.2	&	28,29 \\
1919+0021	&	1	&	50174	&	1.29	&	2 \\ \tableline
1922+2018	&	1	&	n/a	&	n/a	&	1 \\ \tableline
1932+2220	&	4	&	46900	&	4450	&	29 \\
	&			&	50264	&	4457	&	2 \\
	&			&	52210	&	12	&	29 \\
	&			&	52394	&	12	&	29 \\ \tableline
1946+2611	&	1	&	53326	&	70	&	29 \\ \tableline
1952+3252	&	5	&	n/a	&	n/a	&	28 \\
	&			&	51967	&	2.25	&	31 \\
	&			&	52385	&	0.72	&	31 \\
	&			&	52912	&	1.29	&	31 \\
	&			&	53305	&	0.51	&	31 \\ \tableline
2021+3651	&	1	&	52630.07	&	2587	&	18 \\ \tableline
2040+1657	&	1	&	53142	&	n/a	&	29 \\ \tableline
2116+1414	&	3	&	47972	&	0.2	&	28,29 \\
	&			&	49950	&	0.07	&	29 \\
	&			&	51357	&	0.11	&	29 \\ \tableline
2225+6535	&	4	&	43072	&	1707	&	1 \\
	&			&	51900	&	0.14	&	31 \\
	&			&	52950	&	0.08	&	31 \\
	&			&	53434	&	0.19	&	31 \\ \tableline
2229+6114	&	1	&	53064	&	1133	&	29 \\ \tableline
2257+5909	&	1	&	49463.2	&	0.92	&	2 \\ \tableline
2301+5852	&	1	&	52443.9	&	4100	&	19 \\ \tableline
2330$-$2005	&	1	&	n/a	&	n/a	&	1 \\ \tableline
2337+6151	&	1	&	53639	&	20000	&	29 \\ \tableline
\enddata
\tabletypesize{\normal}
\tablerefs{[1] \citet{lyn00}, [2] \citet{kra03} [3] \citet{wan00}, [4] \citet{won01}, [5] \citet{she96}, [6] \citet{morii05},
[7] \citet{mar04},
[8] \citet{zhang01}, [9] \citet{Camilo00}, [10] \citet{wan00},
[11] \citet{camilo04}, [12] \citet{manchester74}, [13] \citet{torii00}, [14] \citet{kaspi00}, [15] \citet{kas03},
[16] \citet{hobbs02}, [17] \citet{cognard04}, [18] \citet{hessels04},
[19] \citet{kaspi03b}, [20] \citet{cordes88}, [21] \citet{shabanova98}, [22] \citet{shabanova05}, [23] \citet{lohsen75},
[24] \citet{lohsen81}, [25] \citet{wwmzly01}, [26] \citet{downs81a}, [27] \citet{dml02}, [28] \citet{hobbs04}, 
[29] \citet{kra05}, [30] \citet{dbrlf04} [31] \citet{jan06} [32] \citet{psurv_2004} [33] \citet{mid06} [34] \citet{fb06}}
\tablecomments{n/a: not available}
\label{tab:gli1}
\end{deluxetable}

\begin{deluxetable}{cccclcc}
\tablewidth{0pt}
\tablecaption{Observing time intervals for pulsars with $N_{\rm g}>5$.}
\tablehead{PSR J & $t_{\rm min}$ & $t_{\rm max}$\\
& (MJD) &  (MJD) }

\startdata
0358+5413 &  41807	& 53546	\\
0534+2200&   \phantom{1} 40466 $\ast$	& \phantom{12}53476 $\ast$ \\
0537$-$6910& 51197	& 53952 \\
0631+1036 &  \phantom{1} 50186 $\ast$ & \phantom{12}53621 $\ast$ \\
0835$-$4510& \phantom{1} 40140 $\ast$ & \phantom{12}53960 $\ast$ \\
1341$-$6220 & 47915& 51022 \\
1740$-$3015 &\phantom{1} 46770 $\ast$ & 53190 \\
1801$-$2304& 46697 & \phantom{12}53356 $\ast$ \\
1825$-$0935& \phantom{1} 48300 $\ast$& \phantom{12}52803 $\ast$ \\ \tableline
\enddata
\tabletypesize{\normal}
\tablecomments{$\ast$: Segmented data spans are not published.
$t_{\rm min}$ and $t_{\rm max}$ are estimated
by eye from graphed spin-down histories,
where available, or else from the first and last glitches by default.}
\label{tab:gli1b}
\end{deluxetable}

\begin{table}
\caption{\label{tab:gli2} Power-law size distribution parameters for pulsars
with $N_{\rm g} > 5$.}
\begin{center}
\begin{tabular}{lrrrr} 
\tableline
\tableline
PSR J & $a_-$ & $a$ & $a_+$ & $P_{\rm KS}$ \\ \tableline
0358$+$5413 & 1.5 & 2.4 & 5.2 & 0.9913 \\
0534$+$2200 & 1.1 & 1.2 & 1.4 & 0.86 \\
0537$-$6910 & 0.39 & 0.42 & 0.43 & 0.36 \\
0631$+$1036 & 1.2 & 1.8 & 2.7 & 0.99896 \\
0835$-$4510 & $-0.20$ & $-0.13$ & 0.18 & 0.908 \\
1341$-$6220 & 1.2 & 1.4 & 2.1 & 0.77 \\
1740$-$3015 & 0.98 & 1.1 & 1.3 & 0.9920 \\
1801$-$2304 & 0.092 & 0.57 & 1.1 & 0.99968 \\
1825$-$0935 & $-0.30$ & 0.36 & 1.0 & 0.99904 \\ 
\tableline
\tableline
\end{tabular}
\end{center}
\end{table}

\begin{table}
\caption{\label{tab:gli3}
Two-component size distribution parameters for quasiperiodic glitchers.}
\begin{center}
\begin{tabular}{lrrrrrr}
\tableline
\tableline
PSR J & $a_-$ & $a$ & $a_+$ & 
  $C_{\rm p}$ & $(\Delta\nu/\nu)_{\rm c}$ & $P_{\rm KS}$
\\ \hline
0537$-$6910 & 0.22 & 0.44 & 0.68 & 
  $0.25\pm0.05$ & $(3.0\pm0.5)\times10^{-7}$ & 0.81
\\
0835$-$4510 & $-0.49$ & 0.11 & 0.44 &
  $0.15\pm0.05$ & $(2.5\pm0.5)\times10^{-6}$ & 0.970 \\
\tableline
\tableline
\end{tabular}
\end{center}

\end{table}

\begin{deluxetable}{crrrrrrr}
\rotate
\tablewidth{0pt}
\tablecaption{ \label{tab:gli4}
Poissonian waiting-time distribution parameters for pulsars
with $N_{\rm g} > 5$.}
\tablehead{
PSR J & $\Delta t_{\rm min}^{(<)}$ & $\Delta t_{\rm min}^{(>)}$
  & $\Delta t_{\rm max}$
  & $\lambda_-$ & $\lambda$ & $\lambda_+$ & $P_{\rm KS}$
\\
  & (d) & (d) & (d)
  & $({\rm yr^{-1}})$ & $({\rm yr^{-1}})$ & $({\rm yr^{-1}})$ }
\startdata
0358$+$5413 & 4 & 30 & 11739 & 0.21 & 0.57 & 1.3 & 0.999960
\\
0534$+$2200 & 2 & 18 & 13010 & 0.57 & 0.91 & 1.3 & 0.982
\\
0537$-$6910 & 3 & 19 & 2755 & n/a & 2.6 & n/a & 0.31
\\
0631$+$1036 & 2 & 16 & 3435 & 0.55 & 0.95 & 1.9 & 0.9970
\\
0835$-$4510 & 2 & 24 & 13820 & 0.33 & 0.35 & 0.42 & 0.45
\\
1341$-$6220 & 4 & 260 & 3107 & 1.2 & 1.8 & 5.6 & 0.980
\\
1740$-$3015 & 4 & 100 & 6330 & 1.2 & 1.5 & 2.5 & 0.928
\\
1801$-$2304 & 4 & 200 & 6659 & 0.35 & 0.55 & 0.88 & 0.962
\\
1825$-$0935 & 4 & 16 & 4503 & 0.48 & 0.91 & 1.8 & 0.9989
\\ \tableline
\enddata
\tablecomments{n/a: not applicable, as the best fit yields $P_{\rm KS} < 0.32$.}
\end{deluxetable}

\begin{deluxetable}{lrrrrrrrccc}
\rotate
\tablewidth{0pt}
\tablecaption{\label{tab:gli5}
Two-component waiting-time distribution parameters for quasiperiodic glitchers.}
\tablehead{PSR J & $\lambda_-$ & $\lambda$ & $\lambda_+$
  & $C_{\rm p}'$ & $\Delta t_{\rm c}$ & $P_{\rm KS}$ 
\\
 & $({\rm yr^{-1}})$ & $({\rm yr^{-1}})$ & $({\rm yr^{-1}})$ & 
 & (yr) &} 
\startdata
0537$-$6910 & 2.0 & 2.6 & 3.3 & 
 $0.25\pm0.05$ & $0.3\pm0.1$ & 0.80 
\\
0835$-$4510 & 0.27 & 0.43 & 0.62 & 
 $0.25\pm0.05$ & $2.8\pm0.1$ & 0.968
\\ \tableline
\enddata
\end{deluxetable}

\newpage

\end{document}